\documentclass[amsmath,amssymb,prc,final,superscriptaddress,showpacs,twocolumn]{revtex4} %showkeys

\usepackage{bm,hyphenat,xspace}
\usepackage{graphicx,epsfig}

\newcommand{\scl}{0.63}

\newcommand{\Eq}{Eq.}
\newcommand{\Eqs}{Eqs.}
\newcommand{\Fig}{Fig.}
\newcommand{\Figs}{Figs.}
\newcommand{\Ref}{Ref.}
\newcommand{\Refs}{Refs.}
\newcommand{\Sect}{Sec.}
\renewcommand {\vec}[1]{{\mathbf{#1}}}
\newcommand {\mbf}[1]{{\mathbf{#1}}}

\newcommand {\Kpl}{\vec{K}_{+}}

\newcommand{\fm}{\;\mathrm{fm}}
\newcommand{\cm}{\mathrm{c\!\:\!.m\!\:\!.}}
\newcommand{\He}{{}^3\mathrm{He}}
\newcommand{\Hh}{{}^3\mathrm{H}}
\newcommand{\extcc}{\emph{Coulomb externally corrected}}
\newcommand{\zr}{\mathcal{Z}_R^{-\frac12}}
\newcommand{\zR}{\mathcal{Z}_R}

\begin{document}

\title {Momentum-space treatment of Coulomb interaction in 
three-nucleon reactions with two protons}

\author{A.~Deltuva} 
\email{deltuva@cii.fc.ul.pt}
\thanks{on leave from Institute of Theoretical Physics and Astronomy,
Vilnius University, Vilnius 2600, Lithuania}
\affiliation{Centro de F\'{\i}sica Nuclear da Universidade de Lisboa, 
P-1649-003 Lisboa, Portugal }
\affiliation{Institut f\"ur Theoretische Physik,  Universit\"at Hannover,
  D-30167 Hannover, Germany}

\author{A.~C.~Fonseca} 
\affiliation{Centro de F\'{\i}sica Nuclear da Universidade de Lisboa, 
P-1649-003 Lisboa, Portugal }

\author{P.~U.~Sauer}
\affiliation{Institut f\"ur Theoretische Physik,  Universit\"at Hannover,
  D-30167 Hannover, Germany}
\received{January 31, 2005}

%\keywords{$\mathit{NN}$-interaction, $\mathit{Nd}$-scattering}

\begin{abstract}
The Coulomb interaction between the two protons is included in the calculation
of proton-deuteron elastic scattering, radiative proton-deuteron capture
and two-body electromagnetic disintegration of ${}^3\mathrm{He}$.
The hadron dynamics is based on the purely nucleonic
charge-dependent (CD) Bonn potential and its realistic extension
CD Bonn + $\Delta$ to a coupled-channel two-baryon potential,
allowing for single virtual $\Delta$-isobar excitation.
Calculations are done using integral equations in momentum space.
The screening and renormalization approach is employed for including the
Coulomb interaction.
Convergence of the procedure is found already at moderate screening radii.
The reliability of the method is demonstrated.
The Coulomb effect on observables is seen at low energies for the whole
kinematic regime. In proton-deuteron elastic scattering  at higher energies 
the Coulomb effect is confined to forward scattering angles;
the $\Delta$-isobar effect found previously remains unchanged by Coulomb.
In electromagnetic reactions Coulomb competes with other effects in a
complicated way.
\end{abstract}

 \maketitle

%%%%%%%%%%%%%%%%%%%%%%%%%%%%%%%%%%%%%%%%%%%%%%%%%%%%%%%%%%%%%%%%%%%%%%%%%%%%%%%
\section{Introduction \label{sec:intro}}

Experimentally, hadronic three-nucleon scattering is predominantly studied in
proton-deuteron $(pd)$ reactions, i.e., in $pd$ elastic scattering 
and breakup: Proton and deuteron beams and targets are available,
with and without polarization. The detection of charged particles yields 
complete experiments. 
In contrast, the charge-symmetric neutron-deuteron $(nd)$ reactions 
are much more difficult to perform, since neutron beams are scarce,
neutron targets non existing, and the detection of two neutrons is a
complicated experimental endeavor.
In electromagnetic (e.m.) reactions, proton-deuteron radiative capture
has a corresponding advantage over neutron-deuteron capture and,
furthermore, $\He$ is a safer target with easier detectable breakup
products  compared with $\Hh$.

In contrast, the Coulomb interaction between the two protons is a
nightmare for the theoretical description of three-nucleon reactions.
The Coulomb interaction is well known, in contrast to the strong 
two-nucleon and three-nucleon potentials mainly studied in three-nucleon
scattering. However, due to its $1/r$ behavior, the Coulomb interaction does
not satisfy the mathematical properties required for the formulation
of standard scattering theory.
When the theoretical description of three-particle scattering 
is attempted in integral form, the Coulomb interaction renders the 
standard equations ill-defined; the kernel of the equations is noncompact.
When the theoretical description is based on differential equations,
the asymptotic boundary conditions for the wave function 
have to be numerically imposed on the trial solutions and,
in the presence of the Coulomb interaction,
those boundary conditions are nonstandard.

There is a long history of theoretical prescriptions for the solution of the
Coulomb problem in three-particle scattering, where different procedures
are followed by the groups involved.
A modified momentum-space integral equation approach is used in 
\Refs~\cite{berthold:90a,alt:02a}, whereas
the configuration-space differential equation approach is used
in \Ref~\cite{kievsky:01a} in a variational framework  and 
in \Refs~\cite{chen:01a,suslov:04a} in the framework of the Faddeev equations.
There are more recent formulations \cite{alt:04a,oryu:04a} of exact
scattering equations with Coulomb which, however, have not matured yet
into practical applications.
In addition there exist approximate schemes:
The most brutal one is the description without Coulomb
for the three-nucleon system with two protons
at those energies and in those kinematical
regimes in which the Coulomb interaction is believed to be irrelevant
for observables; such an approximation has become standard in recent
years~\cite{gloeckle:96a}, and, to our own guilt, we admit
 having used it~\cite{deltuva:03a}.
Reference~\cite{doleschall:82a} extends the assumed applicability of that
approximation scheme by the addition of external Coulomb correction terms to 
those non-Coulomb results.

In this paper our treatment of the Coulomb interaction is based on the ideas
proposed in \Ref~\cite{taylor:74a} for two charged particle scattering
and extended in \Ref~\cite{alt:78a} for three-particle scattering.
The Coulomb potential is screened, standard scattering theory for 
short-range potentials is used, and the obtained results are corrected
for the unscreened limit. We rely on 
\Refs~\cite{taylor:74a,alt:78a}
with respect to the mathematical rigor of that procedure.
We constrain this paper to the description of reactions involving the
$pd$ system. Thus, we leave out breakup in $pd$ scattering 
and three-body breakup in e.m. reactions with $\He$.
We explain the features of our procedure in order to ease the understanding
for the uninitiated reader and to point out differences of our
treatment relative to \Refs~\cite{berthold:90a,alt:02a}, which
also are based on \Refs~\cite{taylor:74a,alt:78a}:

(1) The calculations of \Refs~\cite{berthold:90a,alt:02a} 
need improvement with respect to the hadronic interaction. Whereas
\Refs~\cite{berthold:90a,alt:02a} limited themselves to the use of low-rank 
separable potentials, we use modern two-nucleon potentials and three-nucleon
forces in full without separable expansion.
In particular, the results of this paper are based on
the purely nucleonic charge-dependent (CD) Bonn potential~\cite{machleidt:01a} 
and on its  coupled-channel extension CD Bonn + $\Delta$
\cite{deltuva:03c}, allowing for a single  virtual
$\Delta$-isobar excitation and fitted to the experimental data 
with the same degree of accuracy as CD Bonn itself. 
In the three-nucleon system the $\Delta$ isobar mediates an effective 
three-nucleon force and effective two- and three-nucleon currents,
both consistent with the underlying effective two-nucleon force.
A reliable technique \cite{deltuva:03a} for solving the three-particle 
Alt-Grassberger-Sandhas (AGS) equation~\cite{alt:67a} without Coulomb
is at our disposal. We extend that technique to include the screened Coulomb 
potential between the protons. Thus, the form of our three-particle
equations including the screened Coulomb potential is completely different
from the quasiparticle equations of two-body type
solved in \Refs~\cite{berthold:90a,alt:02a}.

\begin{figure}[!]
\begin{center}
\includegraphics[scale=0.5]{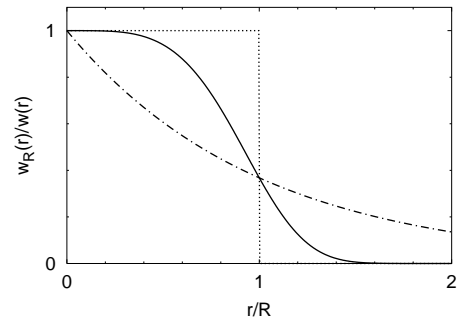}
\end{center}
\caption{\label{fig:wr}
Screening function $w_R(r)/w(r)$ as function of the proton-proton distance 
$r$ for characteristic values of the parameter $n$ in \Eq~\eqref{eq:wr}: 
$n=1$ (dashed-dotted curve) corresponds to Yukawa screening,
$n=4$ (solid curve) is the choice of this 
paper, and $n \to \infty$ (dotted curve) corresponds to a sharp cutoff.}
\end{figure}

(2) We work with a Coulomb potential $w_R$, screened around the
separation $r=R$ between two charged baryons.
We choose $w_R$ in configuration space as 
\begin{gather} \label{eq:wr}
w_R(r) = w(r) \; e^{-(r/R)^n},
\end{gather}
with the true Coulomb potential $w(r) = \frac{\alpha}{r}$,
$\alpha$ being the fine structure constant and
$n$ controlling the smoothness of the screening. 
We prefer to work with a sharper screening than the Yukawa screening
$(n=1)$ of \Refs~\cite{berthold:90a,alt:02a}. We want to ensure that the 
screened Coulomb potential $w_R$ approximates well the true Coulomb one
$w$ for distances $r<R$  and simultaneously vanishes rapidly for $r>R$, 
providing a comparatively fast convergence of the partial-wave expansion.
 In contrast, the sharp cutoff  $(n \to \infty)$
yields an unpleasant oscillatory behavior in the momentum-space representation,
leading to convergence problems. 
We find values $3 \le n \le 6$ to provide a sufficiently smooth, 
but at the same time a sufficiently rapid screening around $r=R$; 
$n=4$ is our choice for the results of this paper, unless indicated otherwise. 
The screening functions for different $n$ values are compared
in \Fig~\ref{fig:wr}.

(3) The screening radius $R$ is chosen much larger than the range of the
strong interaction which is of the order of the pion wavelength
$\hbar/m_\pi c \approx 1.4\fm$. Nevertheless, 
the screened Coulomb potential $w_R$ is of short range in the sense
of scattering theory. Standard scattering theory is therefore applicable.
However, the partial-wave expansion of the pair interaction requires
much higher angular momenta than the one of the strong two-nucleon potential
alone.

(4) The screening radius $R$ will always remain very small compared with
the nuclear screening distances which are of atomic scale, i.e., 
$10^5\fm$. Thus, the employed screened Coulomb potential $w_R$ 
is unable to simulate the physics of nuclear screening properly 
and even more all features of the true Coulomb potential. Therefore 
$w_R$ is unable to yield the Coulomb scattering amplitude as well as
the logarithmic distortion of the Coulomb wave function and,
consequently, the true Coulomb phase shifts.
However, since the Coulomb scattering amplitude and the Coulomb phase shifts
are known and since their occurrence in the three-particle scattering 
amplitudes can be spotted, approximate calculations with screened Coulomb 
$w_R$ can be corrected for their shortcomings in a controlled way.
References~\cite{taylor:74a,alt:78a} 
give the prescription for the correction procedure which we follow here,
and that involves the renormalization of the on-shell amplitudes
in order to get the proper unscreened Coulomb limit.

(5) After the indicated corrections (4), the predictions for observables
of three-nucleon reactions have to show independence from the choice
of the screening radius $R$, provided it is chosen sufficiently large.
That convergence will be our internal criterion for the reliability
of our Coulomb treatment.

Section~\ref{sec:th} describes the practical working of the above 
approximation program in detail.
Section~\ref{sec:res} presents some characteristic effects of Coulomb
in three-nucleon reactions.
Section~\ref{sec:concl} gives our conclusions.

\section{Treatment of Coulomb interaction between protons
 \label{sec:th}}

Section~\ref{sec:intro} recalled the general idea for including the 
Coulomb interaction in $pd$ scattering 
and in related e.m. reactions by screening and renormalization.
This section provides the theoretical framework on which we base
our practical procedure. 
We are aware that the equations given here have been developed
in \Ref~\cite{alt:78a}, but their practical realization in
\Refs~\cite{berthold:90a,alt:02a} differs substantially from the work
presented here. For completeness we rederive some of the equations
and explain how we solve them; our description aims at
elastic $pd$ scattering and e.m. reactions involving the $pd$ system.
However, the essence of the procedure can already be well seen in 
proton-proton $(pp)$ scattering which we therefore use as 
an illustrative example. The example is
also an especially useful test, since exact results for the inclusion
of the Coulomb interaction are readily available; we recover the exact
results, easily obtainable in calculations with a sharp cutoff 
Coulomb~\cite{machleidt:01a}.

The numerical results presented in this section 
refer to the coupled-channel potential CD Bonn + $\Delta$
which allows for single $\Delta$-isobar excitation. Coulomb acts
between the two protons and between the proton and the $\Delta^+$ in the
coupled channel with an isobar.

\subsection{Proton-proton scattering \label{sec:thpp}}
 
The two protons interact through the strong potential $v$ and the 
Coulomb potential $w$.
We introduce the full resolvent $g^{(R)}$ for the auxiliary situation
in which the Coulomb potential $w$ is replaced by the screened potential
$w_R$
\begin{subequations}
  \begin{gather}
    g^{(R)}(z) = (z - h_0 - v - w_R)^{-1},
  \end{gather}
  where $h_0$ is the kinetic energy operator.
  The full resolvent $g^{(R)}(z)$
  yields the full scattering state when acting on a plane-wave state
  $|\mbf{p} \nu \rangle $ of relative momentum $\mbf{p}$, energy $e(p)$,
  and discrete two-particle quantum numbers $\nu$ and taking the appropriate
  limit $z = e(p) + i0$.
  The full resolvent therefore also yields the desired $S$ matrix.
  The full resolvent $g^{(R)}(z)$ 
  depends on the  screening radius $R$ for Coulomb and that dependence is 
  notationally indicated. Next, we discuss formal manipulations of the full
  resolvent. It can be decomposed according to
  \begin{gather}
    g^{(R)}(z) = g_0(z) + g_0(z) t^{(R)}(z) g_0(z)
  \end{gather}
\end{subequations}
with the free resolvent
\begin{gather}
  g_0(z) = (z - h_0)^{-1}
\end{gather}
and the full transition matrix 
\begin{gather} \label{eq:tR}
  t^{(R)}(z) = (v+w_R) + (v+w_R) g_0(z) t^{(R)}(z).
\end{gather}
Of course, $ t^{(R)}(z) $ must contain the pure Coulomb transition
matrix $t_R(z)$ derived from the screened Coulomb potential alone
\begin{gather} \label{eq:t_R}
  t_R(z) = w_R + w_R g_0(z) t_R(z),
\end{gather}
which may be isolated in the full resolvent
\begin{gather} \label{eq:gr2}
  \begin{split}
  g^{(R)}(z) = {} & g_0(z) + g_0(z) t_R(z) g_0(z)  \\ & + 
  g_0(z) [t^{(R)}(z) - t_R(z)] g_0(z).
  \end{split}
\end{gather}

However, an alternative decomposition of the full resolvent appears 
conceptually neater. Instead of correlating the plane-wave state
$|\mbf{p} \nu \rangle $ in a single step to the full scattering state 
by $ g^{(R)}(z) $, it may be correlated first to a screened Coulomb state
by the screened Coulomb potential $w_R$ through
\begin{subequations}
  \begin{align}
    g_{R}(z) = {}& (z - h_0 - w_R)^{-1}, \\
    g_{R}(z) = {}& g_0(z) + g_0(z) t_R(z) g_0(z).
  \end{align}
\end{subequations}
Thus, the full resolvent can alternatively be decomposed into 
\begin{subequations}
  \begin{align}
    g^{(R)}(z) = {}& g_{R}(z) +  g_{R}(z) \tilde{t}^{(R)}(z) g_{R}(z), \\
    \label{eq:gr3b}
    g^{(R)}(z) = {}& g_0(z) + g_0(z) t_R(z) g_0(z) \nonumber \\ & 
    +g_0(z)\{ [1 + t_R(z) g_0(z)] \tilde{t}^{(R)}(z) \nonumber \\ & \times
    [1 + g_0(z) t_R(z)] \}g_0(z)
  \end{align}
\end{subequations}
with the short-range operator 
\begin{gather} \label{eq:ttr}
 \tilde{t}^{(R)}(z) = v + v g_{R}(z)  \tilde{t}^{(R)}(z).
\end{gather}
Equation~\eqref{eq:gr3b} gives an alternative form for the difference
of transition matrices $[t^{(R)}(z) - t_R(z)]$ in \Eq~\eqref{eq:gr2}, i.e.,
\begin{gather} \label{eq:ttr-}
  t^{(R)}(z) - t_R(z) = [1 + t_R(z) g_0(z)] 
 \tilde{t}^{(R)}(z) [1 + g_0(z) t_R(z)].
\end{gather}
The above equation is the well-known two-potential formula that 
 achieves a clean separation of the full transition
matrix $t^{(R)}(z)$ into a long-range part $t_R(z)$ and a 
short-range part $[t^{(R)}(z) - t_R(z)]$. In this paper
the left-hand side of \Eq~\eqref{eq:ttr-} is calculated directly 
from the potentials $v$ and $w_R$ according to
\Eqs~\eqref{eq:tR} and \eqref{eq:t_R}.
Equation ~\eqref{eq:ttr-} is only introduced by us in order to demonstrate that
$[t^{(R)}(z) - t_R(z)]$, even in the infinite $R$ limit,
is a short-range operator due to the short-range 
nature of $v$ and $\tilde{t}^{(R)}(z)$. However, on-shell,
it is externally distorted due to the screened Coulomb wave generated 
by $[1 + g_0(z) t_R(z)] $ which together with the long-range part $t_R(z)$
does not have a proper limit as $R \to \infty$.
This difficulty brings about the concept of renormalization of
on-shell matrix elements of the operators as proposed in 
\Refs~\cite{taylor:74a,alt:78a} in order to recover the proper results
in the unscreened Coulomb limit.

According to \Refs~\cite{taylor:74a,alt:78a}, the $pp$ transition 
amplitude  $\langle \mbf{p}_f \nu_f |t |\mbf{p}_i \nu_i \rangle$, referring
to the strong potential $v$ and the unscreened Coulomb potential $w$,
%and allowing the calculation of scattering observables in the standard way,
is obtained via the renormalization of the on-shell $t^{(R)}(z)$ 
with $z = e(p_i)+i0$ in the infinite $R$ limit
\begin{subequations}\label{eq:tC}
\begin{gather} \label{eq:tC1}
 \begin{split}
 \langle \mbf{p}_f \nu_f |t |\mbf{p}_i \nu_i \rangle = {} &
  \lim_{R \to \infty} \{ \zr(p_f) \langle \mbf{p}_f \nu_f | \\ & \times
  t^{(R)}(e(p_i)+i0) |\mbf{p}_i \nu_i \rangle \zr(p_i) \}.
 \end{split}
\end{gather}
The transition amplitude $\langle \mbf{p}_f \nu_f |t |\mbf{p}_i \nu_i \rangle$
 connects the initial and final states
$|\mbf{p}_i \nu_i \rangle $ and $|\mbf{p}_f \nu_f \rangle $,
$p_f=p_i$, of the considered reaction.
However, \Eq~\eqref{eq:tC1} as it stands is not suitable for the
numerical calculation of the full transition amplitude; instead,
the split of the full transition matrix $t^{(R)}(z)$ into 
long- and short-range parts is most convenient.
For the on-shell screened Coulomb transition matrix $t_R(z)$, 
contained in $t^{(R)}(z)$, the limit in \Eq~\eqref{eq:tC1} 
can be carried out  analytically, yielding 
the true Coulomb transition amplitude 
$\langle \mbf{p}_f \nu_f | t_C |\mbf{p}_i \nu_i \rangle$ 
\cite{taylor:74a}, i.e.,
\begin{gather} \label{eq:tC2}
\begin{split}
\langle \mbf{p}_f \nu_f |t |\mbf{p}_i \nu_i \rangle = {} &
 \langle \mbf{p}_f \nu_f |t_C |\mbf{p}_i \nu_i \rangle  \\ &
 +   \lim_{R \to \infty} \{ \zr(p_f) 
\langle \mbf{p}_f \nu_f | [t^{(R)}(e(p_i)+i0) \\ &  
- t_R(e(p_i)+i0)] |\mbf{p}_i \nu_i \rangle \zr(p_i) \},
\end{split}
\end{gather}
\end{subequations}
whereas the limit for the remaining short-range part $[ t^{(R)}(z) - t_R(z)]$
of the transition matrix $t^{(R)}(z)$ has to be performed numerically,
but it is  reached with sufficient accuracy at 
finite screening radii $R$. In contrast to 
$\langle \mbf{p}_f \nu_f | t_C |\mbf{p}_i \nu_i \rangle$,
the short-range part $[ t^{(R)}(z) - t_R(z)]$ 
can be calculated using a partial-wave expansion of
\Eqs~\eqref{eq:tR} and \eqref{eq:t_R}.

The renormalization factor for  $R \to \infty $ is a diverging
phase factor  
\begin{subequations} \label{eq:zrp}
  \begin{gather}
    \zR(p) = e^{-2i \phi_R(p)},
  \end{gather}
  where $\phi_R(p)$, though  independent  of the $pp$ relative orbital angular 
  momentum $L$ in the infinite $R$ limit, is given by \cite{taylor:74a}
  \begin{gather}    \label{eq:phiRL}
    \phi_R(p) = \sigma_L(p) -\eta_{LR}(p),
  \end{gather}
  with the diverging screened Coulomb phase shift $\eta_{LR}(p)$ 
  corresponding to standard boundary conditions
  and the proper Coulomb one $\sigma_L(p)$ referring to the 
  logarithmically distorted proper Coulomb boundary conditions.
The form \eqref{eq:phiRL} of the renormalization phase is readily 
understood by looking back to \Eq~\eqref{eq:ttr-} and realizing 
that the external distortion generated by the screened Coulomb wave function
$[1 + g_0(e(p)+i0) t_R(e(p)+i0)] |\mbf{p} \nu \rangle$ carries, 
in each partial wave, the overall phase factor $e^{i \eta_{LR}(p)}$
\cite{rodberg:67a}.
 Except for this overall phase factor, the screened
Coulomb wave approximates well the unscreened one
in the range required by the operator $\tilde{t}^{(R)}(z)$ in 
\Eq~\eqref{eq:ttr-} for distances $r < R$.
Therefore, through the renormalization, that unwanted phase factor is changed
to the  appropriate phase factor $e^{i \sigma_L(p)}$ for the unscreened
Coulomb wave.

  For the screened Coulomb potential of \Eq~\eqref{eq:wr}
  the infinite $R$ limit of $\phi_R(p)$ is known analytically \cite{taylor:74a}
  \begin{gather} \label{eq:phiRln}
    \phi_R(p) =  \kappa(p)[\ln{(2pR)} - C/n] ,
  \end{gather}
\end{subequations}
$\kappa(p) = \alpha \mu/p$ being the Coulomb parameter, 
$\mu$ the reduced $pp$ mass, $C \approx 0.5772156649$ the Euler number,
and $n$ the exponent in \Eq~\eqref{eq:wr}.
The renormalization phase $\phi_R(p)$ to be used in the
actual calculations with finite screening radii $R$ is not unique,
since only the infinite $R$ limit matters,
but the converged results have to show independence of
the chosen form of $\phi_R(p)$.
According to our investigations this is indeed so.
The results presented in this paper are based on the
partial-wave dependent form \eqref{eq:phiRL} of the renormalization factor
for which we find the convergence with $R$ to be slightly faster
than for \eqref{eq:phiRln}.

We  refer to \Refs~\cite{taylor:74a,alt:78a} 
for a rigorous justification of the renormalization procedure of
\Eqs~\eqref{eq:tC} and \eqref{eq:zrp} and proceed here to study the 
numerical convergence of our predictions with increasing 
screening radius $R$ as a practical justification for the validity
of the chosen Coulomb treatment.

\begin{figure}[!]
\begin{center}
\includegraphics[scale=\scl]{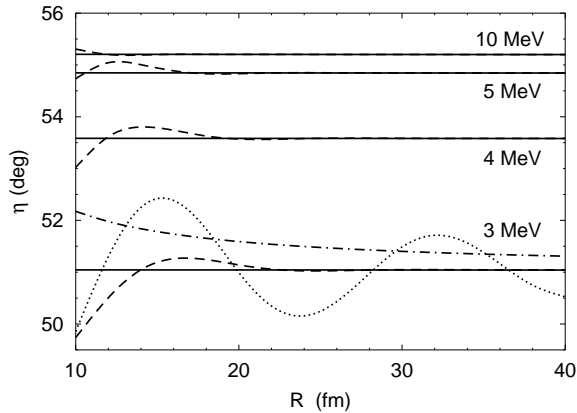}
\end{center}
\caption{\label{fig:pp1S0}
Convergence of the
${}^1S_0$ $pp$ phase shift $\eta$ with screening radius $R$
for proton lab energies  3, 4, 5, and 10~MeV.
Our results derived from \Eq~\eqref{eq:tC2} and given by 
dashed curves are compared with exact results given by solid lines.
At 3~MeV also the results obtained with Yukawa screening (dashed-dotted curve)
and with a sharp cutoff (dotted curve) are shown, demonstrating the
superiority of the screening function chosen in this paper.}
\end{figure}

\begin{figure}[!]
\begin{center}
\includegraphics[scale=\scl]{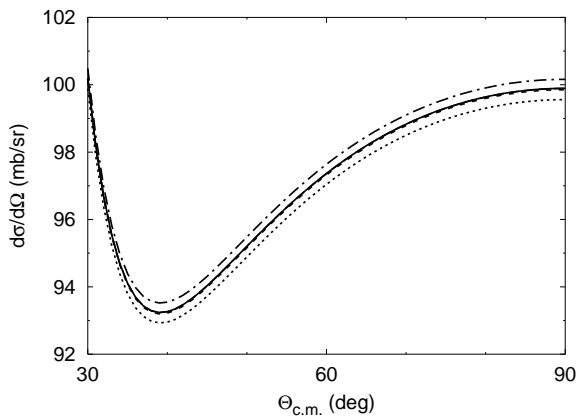}
\end{center}
\caption{\label{fig:ppR}
Convergence of the
differential cross section for $pp$ scattering at 5~MeV proton
lab energy  with screening radius $R$.
The cross section is shown
as function of the c.m. scattering angle. Exact results given
by the solid curve are compared to results with screening radius 
$R = 10$, 15 and 20~fm, given by dotted, dashed-dotted and dashed 
 curves, respectively. Results obtained with $R > 20\fm$ are not
distinguishable from the exact results.}
\end{figure}

\begin{figure}[!]
\begin{center}
\includegraphics[scale=\scl]{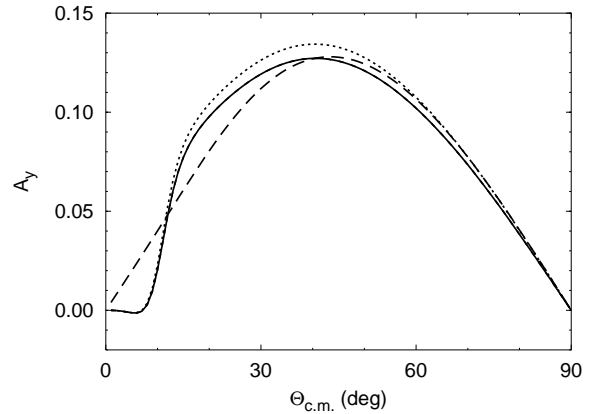}
\end{center}
\caption{\label{fig:ppcmp}
Proton analyzing power for $pp$ scattering at 100~MeV proton lab energy
as function of the c.m. scattering angle. Exact results, given
by the solid curve, are indistinguishable from the results of our Coulomb 
treatment with screening radius $R \ge 10\fm$; 
the dashed curve corresponds to \emph{no Coulomb} results,
while the dotted curve to the \extcc{} approximation. }
\end{figure}

The above discussion left out the identity of the two protons. 
Taking the identity of the protons into account,
the transition amplitude $\langle \mbf{p}_f \nu_f |t |\mbf{p}_i \nu_i \rangle$
 of \Eq~\eqref{eq:tC2} has to be calculated for antisymmetrized states.
Practical results based on \Eq~\eqref{eq:tC2} 
are shown in \Figs~\ref{fig:pp1S0} -- \ref{fig:ppcmp}.

The Coulomb effect on the hadronic $pp$ phase shifts $\eta$ is most 
important in the ${}^1S_0$ partial wave.
The convergence with $R$ for the ${}^1S_0$ phase shift,
shown in \Fig~\ref{fig:pp1S0}, is impressive.
The convergence is faster at higher energies.
A screening radius of $R = 20\fm$ (10~fm) suffices for an agreement
within 0.01~deg with the exact phase shift values at all energies above
5~MeV (25~MeV). In contrast,
in order to reproduce the ${}^1S_0$ $pp$ scattering  length
$a^C_{pp} = -7.815\fm$ and the effective range $r^C_{pp} = 2.773\fm$ 
within $0.010\fm$, screening radii larger than $R=100\fm$ are required. 
In comparison to the screening function adopted in this paper,
\Fig~\ref{fig:pp1S0} also proves the convergence with $R$ to be 
rather slow for the Yukawa screening 
and to be of unpleasant oscillatory behavior for a sharp cutoff.

Figure~\ref{fig:ppR} studies the convergence of the result for the 
spin-averaged $pp$ differential cross section  at 5 MeV 
proton lab energy with increasing screening radius $R$.
The screening radius  $R=20\fm$ appears to be
sufficiently large for that energy, since, according 
to \Fig~\ref{fig:ppR}, the results for $R > 20\fm$
are indistinguishable from the exact Coulomb results,
despite the rather fine scale of the plot.
The rate of convergence seen in \Fig~\ref{fig:ppR} is characteristic
for all studied observables at that energy. The convergence of
observables with $R$ is also faster at higher energies; beyond 25 MeV
the radius $R=10\fm$ is amply enough.

Figure~\ref{fig:ppcmp} shows 
the proton analyzing power results for $pp$ scattering at 100~MeV 
proton lab energy. The results are converged with respect to screening
and the exact results are compared with two  approximations,
labeled \emph{no Coulomb} and \extcc{}:
In the \emph{no Coulomb} approximation, 
the Coulomb interaction is omitted completely; 
in the \extcc{} approximation, the Coulomb
scattering amplitude is added to the no-Coulomb one, the latter being
modified for the external Coulomb distortion by multiplication with the 
Coulomb phase factors $e^{i \sigma_L(p)}$ in the initial and final states
\cite{doleschall:82a}.
Whereas the results converged with respect to screening and the exact results
are indistinguishable in \Fig~\ref{fig:ppcmp}, the  approximations
\emph {no Coulomb} and \extcc{} are  pretty poor even at 100~MeV.
Note that for the observable of \Fig~\ref{fig:ppR} the results for both 
approximations lie out of the scale of that plot.

The results presented in \Figs~\ref{fig:ppR} and \ref{fig:ppcmp} are
characteristic for all observables studied. 
We conclude that the employed method for the inclusion of the Coulomb
interaction in $pp$ scattering works satisfactorily.
We see convergence with increasing screening radius $R$ at moderate values.
The convergence in $R$ is more rapid for higher scattering energies;
$R=10\fm$ is sufficient for  proton lab energies above 25~MeV, 
whereas the screening radius is to be increased beyond 20~fm for energies
below 5~MeV. We also note that
the convergence in $R$ is considerably slower for Yukawa screening and 
is of oscillatory behavior for a sharp cutoff.
The exact Coulomb results are correctly approached by the employed
method with satisfactory accuracy, unlike the
\emph{no Coulomb} or the \extcc{} approximations.
The method we use, based on the ideas of 
\Refs~\cite{taylor:74a,alt:78a},
encourages us to carry it over to elastic $pd$ scattering
as \Refs~\cite{berthold:90a,alt:02a} did and to e.m. reactions involving the
$pd$ system.

\subsection{Elastic proton-deuteron scattering \label{sec:thpde}}

This section carries over the treatment of the Coulomb interaction, 
given in \Sect~\ref{sec:thpp} for $pp$ scattering,
to elastic $pd$ scattering. It establishes a theoretical procedure
leading to a calculational scheme.

Each pair of nucleons $(\beta \gamma)$ interacts through the strong 
coupled-channel potential $v_\alpha$ and the Coulomb potential $w_\alpha$. 
We assume that $w_\alpha$ acts formally between all pairs $(\beta \gamma)$
of particles, but it is nonzero only for states with two-charged baryons,
i.e., $pp$ and $p\Delta^+$ states. 
We introduce the full resolvent $G^{(R)}(Z)$ for the auxiliary situation
in which the Coulomb potential $w_\alpha$ is screened with a
screening radius $R$, $w_\alpha$ being replaced by $w_{\alpha R}$, 
\begin{gather} \label{eq:GR1}
G^{(R)}(Z) = (Z - H_0 - \sum_\sigma v_\sigma - \sum_\sigma w_{\sigma R})^{-1},
\end{gather}
where $H_0$ is the three-particle kinetic energy operator. The full resolvent
yields the full $pd$ scattering state when acting on the
channel state $ |\phi_\alpha (\mbf{q}) \nu_\alpha \rangle $ of
relative $pd$ momentum $\mbf{q}$, energy $E_\alpha(q)$ 
and additional discrete quantum numbers $\nu_\alpha$ and  taking the
appropriate limit $Z = E_\alpha(q) + i0$. The full resolvent
therefore also yields the desired $S$ matrix. 
The full resolvent $G^{(R)}(Z)$ 
depends on the  screening radius $R$ for Coulomb and that dependence is 
  notationally indicated; the same will be done for operators related
to $G^{(R)}(Z)$. 
The full resolvent $G^{(R)}(Z)$, following standard 
AGS notation~\cite{alt:67a} of three-particle scattering, may be decomposed
into channel resolvents
\begin{gather} \label{eq:GRa}
 G^{(R)}_\alpha (Z) = (Z - H_0 - v_\alpha - w_{\alpha R})^{-1},
\end{gather}
where, in $pd$ channels $\alpha$,  $w_{\alpha R} = 0$,
and into the full multichannel three-particle transition matrix 
$U^{(R)}_{\beta \alpha}(Z)$ according to 
\begin{gather} \label{eq:GR2}
  G^{(R)}(Z) = \delta_{\beta \alpha}  G^{(R)}_\alpha (Z) +
  G^{(R)}_\beta (Z)  U^{(R)}_{\beta \alpha}(Z)   G^{(R)}_\alpha (Z).
\end{gather}
The full multichannel transition matrix satisfies the 
AGS equation~\cite{alt:67a}
\begin{subequations}\label{eq:UbaT}
  \begin{align} \label{eq:Uba}
     U^{(R)}_{\beta \alpha}(Z) = {} & \bar{\delta}_{\beta \alpha} G_0^{-1}(Z)
     + \sum_{\sigma} \bar{\delta}_{\beta \sigma} T^{(R)}_\sigma (Z) G_0(Z) 
     U^{(R)}_{\sigma \alpha}(Z), 
  \end{align}
 where the two-particle transition matrix is derived from the full
channel interaction $v_\alpha + w_{\alpha R}$, i.e.,
 \begin{align} \label{eq:TR}
   T^{(R)}_\alpha (Z) = {}& (v_\alpha + w_{\alpha R}) + 
     (v_\alpha + w_{\alpha R})  G_0(Z) T^{(R)}_\alpha (Z),
  \end{align}
\end{subequations}
$G_0(Z) = (Z - H_0)^{-1}$ being the free resolvent and 
$\bar{\delta}_{\beta \alpha} = 1 - {\delta}_{\beta \alpha}$.
Of course, the full multichannel transition matrix
 $U^{(R)}_{\beta \alpha}(Z)$ must contain the pure Coulomb transition 
matrix $T^{\cm}_{\alpha R} (Z)$ derived from the screened Coulomb 
potential $W^{\cm}_{\alpha R}$ between the spectator proton
and the center of mass (c.m.) of the remaining neutron-proton $(np)$ pair
 in channel $\alpha$,
\begin{gather} \label{eq:Tcm}
T^{\cm}_{\alpha R} (Z) = W^{\cm}_{\alpha R} + 
W^{\cm}_{\alpha R} G^{(R)}_{\alpha} (Z) T^{\cm}_{\alpha R} (Z),
\end{gather}
the $pd$ channel being one of those channels $\alpha$.
This part may therefore be isolated in the full resolvent according to
\begin{gather}
  \begin{split}
    G^{(R)}(Z) = {} & {\delta}_{\beta \alpha}  G^{(R)}_{\alpha} (Z) +
G^{(R)}_{\beta}(Z) \delta_{\beta\alpha} 
T^{\cm}_{\alpha R}(Z) G^{(R)}_{\alpha}(Z) \\ & 
+ G^{(R)}_{\beta}(Z) [ U^{(R)}_{\beta \alpha}(Z) -
\delta_{\beta\alpha} T^{\cm}_{\alpha R}(Z)] G^{(R)}_{\alpha}(Z).
\label{eq:GR3}
\end{split}
\end{gather}

Nevertheless, as we have done in \Sect~\ref{sec:thpp},
an alternative decomposition of the full resolvent, which appears
conceptually neater for the purpose of elastic $pd$ scattering,
may be developed based on the following idea.
Instead of correlating the plane-wave channel state
$ |\phi_\alpha (\mbf{q}) \nu_\alpha \rangle $ in a single step to the full
scattering state by $G^{(R)}(Z)$, it may be correlated first to a screened
Coulomb state of proton and deuteron by the screened Coulomb potential
$W^{\cm}_{\alpha R}$ between a proton and the c.m. of an $np$ pair through
\begin{subequations}
  \begin{align}
    G_{\alpha R}(Z) = {}& 
    (Z - H_0 - v_\alpha - w_{\alpha R} - W^{\cm}_{\alpha R})^{-1}, \\
    G_{\alpha R}(Z) = {}& G^{(R)}_{\alpha}(Z) + 
    G^{(R)}_{\alpha}(Z) T^{\cm}_{\alpha R}(Z) G^{(R)}_{\alpha}(Z),
  \end{align}
\end{subequations}
where, in each channel $\alpha$,  $w_{\alpha R}$ and $W^{\cm}_{\alpha R}$ 
are never simultaneously present: When $\alpha$ corresponds to a $pp$ pair,
$w_{\alpha R}$ is present and $W^{\cm}_{\alpha R} = 0$;
when $\alpha$ denotes an $np$ pair, $w_{\alpha R} = 0$
and $W^{\cm}_{\alpha R}$ is present.
Thus, the full resolvent can alternatively be decomposed into
\begin{subequations}
  \begin{align}
    G^{(R)}(Z) = {} & \delta_{\beta \alpha}   G_{\alpha R}(Z) +
    G_{\beta R}(Z) \tilde{U}^{(R)}_{\beta\alpha}(Z) G_{\alpha R}(Z),
 \\ 
  G^{(R)}(Z) = {} & \delta_{\beta \alpha} G^{(R)}_{\alpha}(Z) +
 G^{(R)}_{\beta}(Z) \delta_{\beta \alpha} T^{\cm}_{\alpha R}(Z) 
 G^{(R)}_{\alpha}(Z)
 \nonumber \\ &
 + G^{(R)}_{\beta}(Z) \{ [1 + T^{\cm}_{\beta R}(Z) G^{(R)}_{\beta}(Z)]
 \tilde{U}^{(R)}_{\beta\alpha}(Z) \nonumber \\ & \times
 [1 + G^{(R)}_{\alpha}(Z) T^{\cm}_{\alpha R}(Z)] \} G^{(R)}_{\alpha}(Z),
 \label{eq:GR4b}
  \end{align}
\end{subequations}
where the  operator $ \tilde{U}^{(R)}_{\beta\alpha}(Z)$
may be calculated through the integral equation
\begin{gather}
  \begin{split}
 \tilde{U}^{(R)}_{\beta\alpha}(Z) = {} &
\bar{\delta}_{\beta \alpha} [G_{\alpha R}^{-1}(Z) + v_{\alpha}] +
{\delta}_{\beta \alpha} \mathcal{W}_{\alpha R}  \\ &
+ \sum_\sigma (\bar{\delta}_{\beta \sigma} v_\sigma +
{\delta}_{\beta \sigma} \mathcal{W}_{\beta R})
G_{\sigma R}(Z) \tilde{U}^{(R)}_{\sigma\alpha}(Z),
\label{eq:tU}
\end{split}
\end{gather}
which is driven by the strong potential $v_\alpha$  and the  potential 
of three-nucleon nature
$\mathcal{W}_{\alpha R} = \sum_{\sigma} 
( \bar{\delta}_{\alpha \sigma} w_{\sigma R} - 
\delta_{\alpha\sigma} W^{\cm}_{\sigma R} ) $.
This potential $\mathcal{W}_{\alpha R}$
accounts for the difference between the direct $pp$ Coulomb interaction
and the one that takes place between the proton and the c.m. of the remaining
bound as well as unbound  $np$ pair.
When calculated between on-shell screened $pd$ Coulomb states,
$\tilde{U}^{(R)}_{\beta\alpha}(Z)$ is of
short-range, even in the infinite $R$ limit.
Equation~\eqref{eq:GR4b} gives an alternative form for the difference of the
transition matrices 
$ [U^{(R)}_{\beta \alpha}(Z) - \delta_{\beta\alpha} T^{\cm}_{\alpha R}(Z)]$
in \Eq~\eqref{eq:GR3}, i.e.,
\begin{gather} \label{eq:U-T}
  \begin{split}
    U^{(R)}_{\beta \alpha}(Z) - \delta_{\beta\alpha} T^{\cm}_{\alpha R}(Z)
    = {} & [1 + T^{\cm}_{\beta R}(Z) G^{(R)}_{\beta}(Z)] 
    \tilde{U}^{(R)}_{\beta\alpha}(Z)  \\ & \times
	  [1 + G^{(R)}_{\alpha}(Z) T^{\cm}_{\alpha R}(Z)].
  \end{split}
\end{gather}
Though we calculate that difference directly from
the potentials $v_\alpha$, $w_{\alpha R}$, and $W^{\cm}_{\alpha R}$
through the numerical 
solution of \Eqs~\eqref{eq:UbaT} and \eqref{eq:Tcm}, 
\Eq~\eqref{eq:U-T} demonstrates that for initial and final $pd$ states
$ [U^{(R)}_{\beta \alpha}(Z) - \delta_{\beta\alpha} T^{\cm}_{\alpha R}(Z)]$
is a short-range operator due to the  nature  of 
$ \tilde{U}^{(R)}_{\beta\alpha}(Z)$  as discussed above,
but it is externally distorted
due to the screened Coulomb wave generated by 
$[1 + G^{(R)}_{\alpha}(Z) T^{\cm}_{\alpha R}(Z)]$.
Thus, \Eq~\eqref{eq:U-T} achieves a clean separation of the full
on-shell transition matrix $U^{(R)}_{\beta \alpha}(Z)$ into the long-range
part  $\delta_{\beta\alpha} T^{\cm}_{\alpha R}(Z)$
and the short-range part 
$ [U^{(R)}_{\beta \alpha}(Z) - \delta_{\beta\alpha} T^{\cm}_{\alpha R}(Z)]$.
On-shell, both parts do not have a proper limit as $R \to \infty$. 
They have to get renormalized as the corresponding amplitudes for $pp$ 
scattering in \Sect~\ref{sec:thpp}, 
in order to obtain the results appropriate for the unscreened Coulomb limit. 

According to \Refs~\cite{taylor:74a,alt:78a}, 
the full  $pd$  transition amplitude %$U_{\beta \alpha}$
for initial and final states 
$ |\phi_\alpha (\mbf{q}_i) \nu_{\alpha_i} \rangle $ and
$ |\phi_\beta (\mbf{q}_f) \nu_{\beta_f} \rangle $,
$q_f = q_i$, referring
to the strong potential $v_\alpha$ and the unscreened Coulomb potential 
$w_\alpha$, is obtained via the renormalization of the on-shell
multichannel transition matrix $ U^{(R)}_{\beta \alpha}(Z)$  with
$Z = E_\alpha(q_i) + i0$ in the infinite $R$ limit
\begin{subequations}\label{eq:UC}
  \begin{gather} \label{eq:UC1}
  \begin{split}
    \langle \phi_\beta (\mbf{q}_f) & \nu_{\beta_f} | U_{\beta \alpha}
    |\phi_\alpha (\mbf{q}_i) \nu_{\alpha_i} \rangle  \\ = {}& 
    \lim_{R \to \infty} \{ \zr(q_f) 
    \langle \phi_\beta (\mbf{q}_f) \nu_{\beta_f} | \\ & \times
    U^{(R)}_{\beta \alpha}(E_\alpha(q_i) + i0) 
    |\phi_\alpha (\mbf{q}_i) \nu_{\alpha_i} \rangle \zr(q_i) \}.
  \end{split}
\end{gather}
As for $pp$ scattering, the split of the full on-shell
multichannel transition matrix $ U^{(R)}_{\beta \alpha}(Z)$ 
into long- and short-range parts is most convenient.
For the screened Coulomb transition matrix $T^{\cm}_{\alpha R}(Z)$,
contained in $ U^{(R)}_{\beta \alpha}(Z)$,
the limit in \Eq~\eqref{eq:UC1} can be carried out analytically, yielding 
the proper Coulomb transition amplitude 
$ \langle \phi_\beta (\mbf{q}_f) \nu_{\beta_f} |T^{\cm}_{\alpha C}
|\phi_\alpha (\mbf{q}_i) \nu_{\alpha_i} \rangle $
\cite{taylor:74a,alt:78a}, i.e.,
\begin{gather} \label{eq:UC2}
  \begin{split}
    \langle  \phi_\beta (\mbf{q}_f) & \nu_{\beta_f} | U_{\beta \alpha}
    |\phi_\alpha (\mbf{q}_i) \nu_{\alpha_i} \rangle   \\ = {}& 
    \delta_{\beta \alpha} 
    \langle \phi_\beta (\mbf{q}_f) \nu_{\beta_f} |T^{\cm}_{\alpha C}
    |\phi_\alpha (\mbf{q}_i) \nu_{\alpha_i} \rangle  \\ & +
    \lim_{R \to \infty} \{ \zr(q_f) 
    \langle \phi_\beta (\mbf{q}_f) \nu_{\beta_f} |
	    [ U^{(R)}_{\beta \alpha}(E_\alpha(q_i) + i0)  \\ & -
	      \delta_{\beta\alpha} T^{\cm}_{\alpha R}(E_\alpha(q_i) + i0)] 
	    |\phi_\alpha (\mbf{q}_i) \nu_{\alpha_i} \rangle
	    \zr(q_i) \}.
  \end{split}  
\end{gather}
\end{subequations}
The limit for the remaining part $[ U^{(R)}_{\beta \alpha}(Z) -
    \delta_{\beta\alpha} T^{\cm}_{\alpha R}(Z)]$
of the multichannel transition matrix has to be performed numerically,
but, due to the short-range nature of that part,
it is reached with sufficient accuracy at 
finite screening radii $R$, and furthermore, 
$[ U^{(R)}_{\beta \alpha}(Z) - \delta_{\beta\alpha} T^{\cm}_{\alpha R}(Z)]$
can be calculated using a partial-wave expansion.

In close analogy with $pp$ scattering, 
the renormalization factor for  $R \to \infty $ is a diverging phase factor  
\begin{subequations} \label{eq:zrq}
  \begin{gather}
    \zR(q) = e^{-2i \phi_R(q)},
  \end{gather}
  where $\phi_R(q)$, though independent of the $pd$ relative  
  angular momentum $l$ in the infinite $R$ limit, is 
  \begin{gather}    \label{eq:phiRl}
    \phi_R(q) = \sigma_l(q) -\eta_{lR}(q), 
  \end{gather}
with the diverging screened Coulomb $pd$ phase shift $\eta_{lR}(q)$ 
corresponding to standard boundary conditions
and the proper Coulomb one $\sigma_l(q)$ referring to the 
logarithmically distorted proper Coulomb boundary conditions.
In analogy to $pp$ scattering the form \eqref{eq:phiRl} of the 
renormalization phase is readily understood by looking back to 
 \Eq~\eqref{eq:U-T}.
For the screened Coulomb potential of \Eq~\eqref{eq:wr} 
the infinite $R$ limit of $\phi_R(q)$ is known analytically 
\begin{gather}  \label{eq:phiRlln}
  \phi_R(q) =  \kappa(q)[\ln{(2qR)} - C/n],
  \end{gather}
\end{subequations}
$\kappa(q) = \alpha M/q$ being the $pd$ Coulomb
parameter and $M$ the reduced $pd$ mass.
The form of the renormalization phase $\phi_R(q)$ to be used in the
actual calculations with finite screening radii $R$ is not unique, 
but, like in \Sect~\ref{sec:thpp},
the converged results show independence of
the chosen form of $\phi_R(q)$.
The results presented in this paper are based on the
partial-wave dependent form \eqref{eq:phiRl} of the renormalization factor
for which we find the convergence with $R$ to be slightly faster
than for \eqref{eq:phiRlln}.

Again we refer to \Refs~\cite{taylor:74a,alt:78a} 
for a rigorous justification of the correction procedure of
\Eqs~\eqref{eq:UC} and \eqref{eq:zrq} and proceed here to study the 
numerical convergence of our predictions with increasing $R$.

We choose an isospin description for the three baryons involved in 
three-nucleon scattering.
In the isospin formalism the isospin $T$ of the interacting pair
and the isospin $t$ of the spectator are coupled to the total
isospin $\mathcal{T}$ with the projection $\mathcal{M_T}$.
Due to the hadronic charge dependence together with the screened Coulomb
interaction in $pp$ and $p\Delta^+$ pair states, i.e., 
in states with isospin $|T M_T \rangle = | 1 1 \rangle$,
the two-baryon transition matrix $T^{(R)}_{\alpha}(Z)$ 
becomes an operator coupling total isospin $\mathcal{T} = \frac12$
and $\mathcal{T} = \frac32$ states according to
\begin{gather} \label{eq:Tiso}
\begin{split}
  \langle (T't') \mathcal{T' M'_T} | {} &  T^{(R)}_{\alpha}(Z)
  | (Tt) \mathcal{T M_T} \rangle \\ = {} &
  \delta_{T'T} \delta_{t't} \delta_{\mathcal{M'_T M_T}} 
  \sum_{M_T m_t} \langle T M_T t m_t| \mathcal{T' M_T}\rangle 
\\ & \times   %% (\mathcal{M_T} \!-\!M_T)
\langle T M_T | T^{(R)}_{\alpha}(Z) | TM_T \rangle
  \langle T M_T t m_t | \mathcal{T M_T} \rangle .
\end{split}
\end{gather}
%In the isospin formalism the nucleons are considered identical
Due to the isospin formulation, the nucleons are therefore considered
identical. However, the discussion has left out the identity of nucleons
till now. Instead of the transition amplitude of \Eq~\eqref{eq:UC2} 
we therefore have to use the properly symmetrized form
\begin{subequations}
  \begin{gather}
  \begin{split} \label{eq:Uasyma}
    \langle \phi_\alpha (\mbf{q}_f) & \nu_{\alpha_f} |
    U |\phi_\alpha (\mbf{q}_i) \nu_{\alpha_i} \rangle  \\ = {} &
    \sum_\sigma 
    \langle \phi_\alpha (\mbf{q}_f) \nu_{\alpha_f} |U_{\alpha \sigma}
    | \phi_\sigma (\mbf{q}_i) \nu_{\sigma_i} \rangle, 
  \end{split}
  \end{gather} \vspace{-5mm}
  \begin{gather}
  \begin{split}\label{eq:Uasymb}
    \langle \phi_\alpha (\mbf{q}_f)  & \nu_{\alpha_f} |
    U |\phi_\alpha (\mbf{q}_i) \nu_{\alpha_i} \rangle  \\ = {} & 
    \langle \phi_\alpha (\mbf{q}_f) \nu_{\alpha_f} |
    T^{\cm}_{\alpha C}|\phi_\alpha (\mbf{q}_i) \nu_{\alpha_i} \rangle
    \\ & +  \lim_{R \to \infty} 
    \{ \zr(q_f) \langle \phi_\alpha (\mbf{q}_f) \nu_{\alpha_f}|
       [ U^{(R)}(E_\alpha(q_i) + i0)  \\ & -
	 T^{\cm}_{\alpha R}(E_\alpha(q_i) + i0)] 
       |\phi_\alpha (\mbf{q}_i) \nu_{\alpha_i} \rangle  \zr(q_i) \}
  \end{split}
  \end{gather}
\end{subequations}
with $U^{(R)}(Z) = U^{(R)}_{\alpha \alpha}(Z) +
U^{(R)}_{\alpha \beta}(Z) P_{231} + U^{(R)}_{\alpha \gamma}(Z) P_{312}$
for the calculation of observables, 
$(\alpha \beta \gamma)$ being cyclic and $P_{231}$ and $P_{312}$ being the
two cyclic permutations of $(\alpha \beta \gamma)$.
 $U^{(R)}(Z)$ satisfies the standard
symmetrized form of the integral equation \eqref{eq:Uba}, i.e.,
\begin{gather} \label{eq:UR}
  U^{(R)}(Z) = P G_0^{-1}(Z) + P T^{(R)}_{\alpha}(Z) G_0(Z) U^{(R)}(Z)
\end{gather}
with $P = P_{231} + P_{312}$. 

The practical implementation of the outlined calculational scheme
faces a technical difficulty. We solve \Eq~\eqref{eq:UR}
in a partial-wave basis. The partial-wave expansion of the
screened Coulomb potential converges rather slowly.
The problem does not occur in $pp$ scattering, since
there the partial waves with different two-baryon total angular momentum $I$
are not coupled and the maximal $I$ 
required for $[ t^{(R)}(z) - t_R(z)]$  is determined 
according to \Eqs~\eqref{eq:ttr} and \eqref{eq:ttr-}
by the range of the hadronic potential $v$.
However, in the calculation of $U^{(R)}(Z)$
all two-baryon partial waves are coupled dynamically; the required
maximal $I$ is determined by the range of the screened Coulomb potential
and is considerably higher than required for the hadronic potential alone.
In this context, the perturbation theory for higher two-baryon partial
waves developed in \Ref~\cite{deltuva:03b} 
is a very efficient and reliable technical tool for treating the screened 
Coulomb interaction in high partial waves. Furthermore, in practical 
calculations we split the difference of the transition matrices
in \Eq~\eqref{eq:Uasymb} into two parts with different partial-wave
 convergence properties, 
\begin{gather} \label{eq:U-PTP}
  \begin{split}
     U^{(R)}(Z) - T^{\cm}_{\alpha R}(Z)
    = {} & [ U^{(R)}(Z) - PT_{\alpha R}(Z)P] 
    \\ & - [T^{\cm}_{\alpha R}(Z)-PT_{\alpha R}(Z)P],
  \end{split}
\end{gather}
$T_{\alpha R}(Z)$ being the two-baryon screened Coulomb transition matrix
derived from $w_{\alpha R}$ alone and hidden in $T_\alpha^{(R)}(Z)$ 
according to \Eq~\eqref{eq:TR}. The term
$PT_{\alpha R}(Z)P$ is the remainder of the three-body operator
$U^{(R)}(Z) - PG_0^{-1}(Z)$ in the absence of the strong force, and
it is contained in $U^{(R)}(Z)$ as the most important
Coulomb contribution; the difference $[ U^{(R)}(Z) - PT_{\alpha R}(Z)P]$
converges with respect to included two-baryon states considerably faster than
$U^{(R)}(Z)$ alone.
The term $[T^{\cm}_{\alpha R}(Z)-PT_{\alpha R}(Z)P]$ accounts for the
off c.m. $pd$ screened Coulomb interaction and  converges
rather slowly, but the inclusion of very high partial waves is 
much easier than for  $[ U^{(R)}(Z) - PT_{\alpha R}(Z)P]$.
We vary the dividing line between partial waves included exactly and 
perturbatively in $U^{(R)}(Z)$  as well as angular momentum cutoffs 
for both terms in \Eq~\eqref{eq:U-PTP} in order to test the convergence 
and thereby establish the validity of the procedure.
The problem of high partial waves does not occur in 
\Refs~\cite{berthold:90a,alt:02a}, since the authors use the 
quasiparticle formalism and work with  equations of two-body type 
in which the partial-wave decomposition has to be performed only 
with respect to the relative motion of the spectator particle 
and the correlated pair.
Due to technical limitations, \Refs~\cite{berthold:90a,alt:02a}
use low-rank separable potentials for the hadronic interaction
and approximate the two-proton screened Coulomb transition matrix
by the potential. In contrast, we work with a realistic hadronic
interaction (CD Bonn or CD Bonn + $\Delta$)
 without separable approximation, and we never
approximate the energy-dependent pair transition matrix for screened
Coulomb $T_{\alpha R}(Z)$ by the potential $w_{\alpha R}$.
 
With respect to the partial-wave expansion in the actual calculations of 
this paper, we obtain fully converged results by taking into account
the screened Coulomb interaction in two-baryon partial waves
with pair orbital angular momentum $L \le 13$ for the first term
in \Eq~\eqref{eq:U-PTP} and with $L \le 25$ for the second term;
orbital angular momenta $L \ge 7$ can safely be treated 
perturbatively. The above values refer to the screening radius $R=25\,\fm$;
for smaller screening radii the convergence in orbital angular momentum 
is faster. The hadronic interaction is taken into account
 in two-baryon partial waves with total angular momentum $I \le 5$. 
Both three-baryon total isospin $\mathcal{T} = \frac12$
and $\mathcal{T} = \frac32$ states are included.
The maximal three-baryon total angular momentum $\mathcal{J}$
considered is $\frac{31}{2}$.

\begin{figure}[!]
\begin{center}
\includegraphics[scale=\scl]{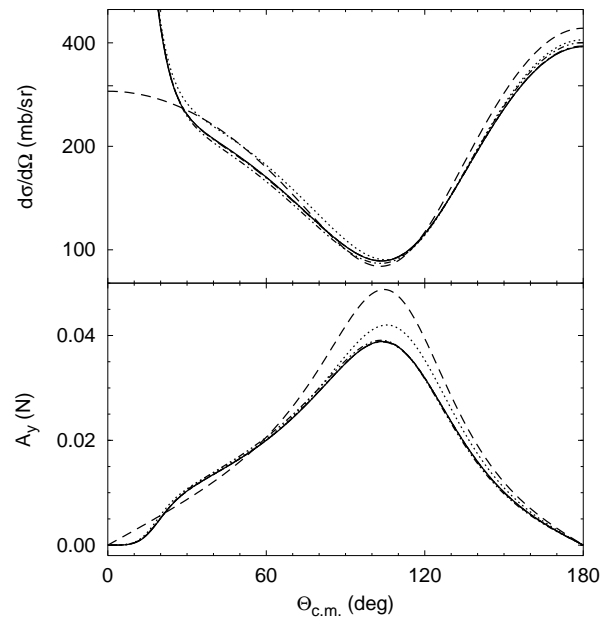}
\end{center}
\caption{\label{fig:pd3R}
Convergence of the differential cross section and of the proton 
analyzing power $A_y(N)$ for $pd$ elastic scattering at 3~MeV 
proton lab energy with screening radius $R$. The observables are
shown as functions of the c.m. scattering angle. 
The hadronic potential is CD Bonn + $\Delta$.
Results obtained with screening radius $R= 5$~fm (dotted curves), 
10~fm (dashed-double-dotted curves),
15~fm (dashed-dotted curves), 20~fm (double-dashed-dotted curves), 
25~fm (solid curves) are compared. Results without Coulomb (dashed curves)
are given as reference for the size of the Coulomb effect.}
\end{figure}

\begin{figure}[!]
\begin{center}
\includegraphics[scale=\scl]{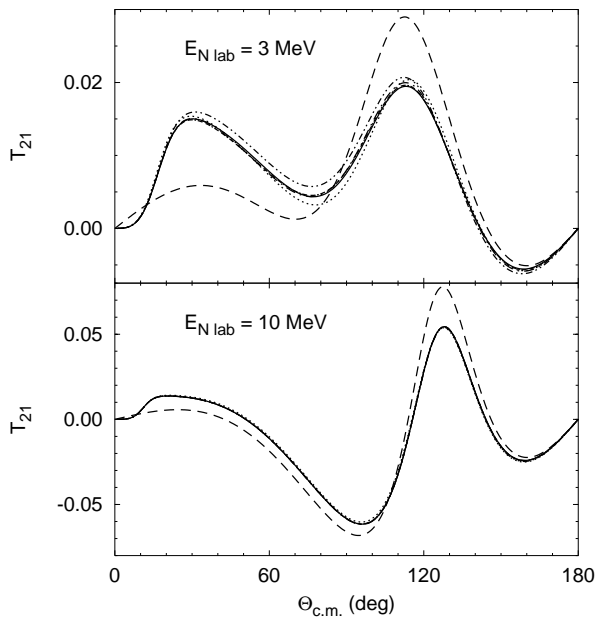}
\end{center}
\caption{\label{fig:T21R}
Convergence of the deuteron tensor analyzing power $T_{21}$ for $pd$ 
elastic scattering at 3 MeV and 10~MeV proton lab energy 
with screening radius $R$. The observable is shown
as function of the c.m. scattering angle. 
The curves are explained in the caption of \Fig~\ref{fig:pd3R}.}
\end{figure}

Figures~\ref{fig:pd3R} and \ref{fig:T21R} study the convergence 
of our method with increasing screening 
radius $R$ according to \Eq~\eqref{eq:Uasymb}.
The comparison with the \emph{no Coulomb} results (dashed curves), 
used till now by us when accounting for $pd$ data, 
gives the size of the Coulomb effect.
First we concentrate on 3~MeV proton lab energy, 
the lowest energy considered in this paper. 
As examples we show the differential cross section, the nucleon analyzing power
$A_y(N)$ which has the most critical convergence behavior according
to \Refs~\cite{berthold:90a,alt:98a},
and the deuteron tensor analyzing power $T_{21}$,
the most slowly converging observable at 3 MeV proton lab energy
according to our experience.
Nevertheless, the convergence is impressive even for those worst cases:
Only $T_{21}$, shown in \Fig~\ref{fig:T21R}, requires a screening radius 
$R > 15\fm$. Convergence is more rapid at higher energies as demonstrated
in \Fig~\ref{fig:T21R} for the deuteron tensor analyzing power $T_{21}$
at 3 and 10~MeV proton lab energy.
The observed convergence strongly suggests the
reliability of the chosen Coulomb treatment. 
Furthermore, the forthcoming 
paper~\cite{deltuva:05b} makes a detailed comparison between the
results obtained by the present technique and those of 
\Ref~\cite{kievsky:01a} obtained from the variational solution
of the three-nucleon Schr\"odinger equation in configuration space
with the inclusion of an \emph{unscreened} Coulomb potential
between the protons and imposing  the proper Coulomb boundary 
conditions explicitly. The agreement, across the board, between the
results derived from two entirely different methods, clearly
indicates that both techniques for including the Coulomb interaction
are reliable; this is another justification for the technique used in this
paper.

\begin{figure}[!]
\begin{center}
\includegraphics[scale=0.58]{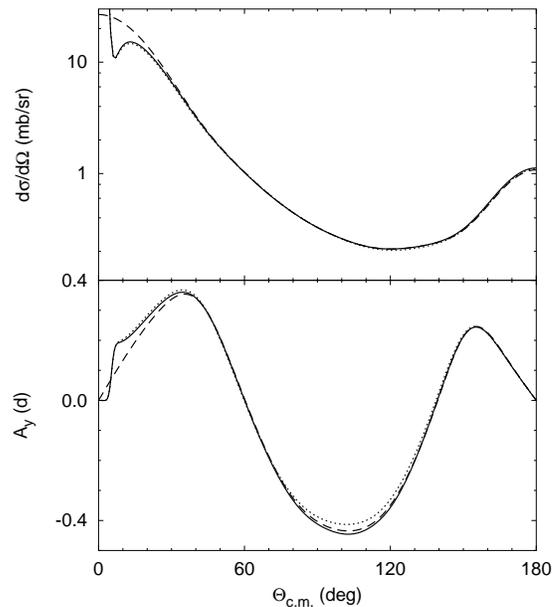}
\end{center}
\caption{\label{fig:pdcmp}
Differential cross section and 
deuteron  analyzing power $A_y(d)$
for $pd$ elastic scattering at 135~MeV proton lab energy
as functions of the c.m. scattering angle. Converged results of the present
Coulomb treatment with $R = 10\fm$ given by solid curves are compared 
to the results  calculated with \emph{no Coulomb} (dashed curves) 
and \extcc{} (dotted curves) approximations.} 
\end{figure}

As in \Fig~\ref{fig:ppcmp} for $pp$ scattering, \Fig~\ref{fig:pdcmp}
compares predictions including the Coulomb interaction with results from
traditional approximate treatments which were labeled before as
\emph{no Coulomb} and \extcc{}.
As already known from \Refs~\cite{berthold:90a,alt:98a}
both approximations are unsatisfactory at low energies.
At higher energies the Coulomb effect is confined more and more
to the forward direction; the \emph{no Coulomb} treatment fails there, 
whereas the \extcc{} approximation is usually not accurate enough 
for larger scattering angles.

The seen Coulomb effects and their physics implications are discussed
in \Sect~\ref{sec:res}.

\subsection{Radiative capture and two-body e.m. disintegration 
of the three-nucleon bound state \label{sec:them}}

For the description of the considered e.m. processes the 
matrix element $\langle \psi^{(-)}_\alpha (\mbf{q}_f) \nu_{\alpha_f} |
  j^{\mu} (\mbf{Q}, \Kpl ) | B \rangle$ 
of the e.m. current operator
between the three-nucleon bound state and the $pd$ scattering state
has to be calculated.
The calculation of that matrix element without Coulomb is discussed in
great length in \Refs~\cite{deltuva:04a,deltuva:04b}.
This subsection only discusses the modification which arises due to
the inclusion of the Coulomb interaction between the charged baryons.
Coulomb is included as a screened potential and the dependence of the
bound and scattering states, i.e., $| B^{(R)} \rangle$  and
$ |\psi^{(\pm)(R)}_\alpha (\mbf{q}_f) \nu_{\alpha_f} \rangle $,
on the screening radius $R$ is notationally made explicit.
In analogy to $pd$ scattering,
the current matrix element referring to the unscreened Coulomb potential
is obtained via renormalization of the matrix element 
referring to the screened Coulomb potential in the infinite $R$ limit
\begin{gather} \label{eq:jR}
  \begin{split}
    \langle \psi^{(-)}_\alpha & (\mbf{q}_f)  \nu_{\alpha_f} |
    j^{\mu} (\mbf{Q}, \Kpl ) | B \rangle \\ = {} & 
    \lim_{R \to \infty} 
    \zr(q_f) \langle \psi^{(-)(R)}_\alpha (\mbf{q}_f) \nu_{\alpha_f} |
    j^{\mu} (\mbf{Q}, \Kpl ) | B^{(R)} \rangle.
  \end{split}
\end{gather}
As for $pd$ scattering, the
practical results presented in this paper are based on the
partial-wave dependent form of the renormalization factor \eqref{eq:phiRl}.
Due to the short-range nature of 
$ j^{\mu} (\mbf{Q}, \Kpl ) | B^{(R)} \rangle$ 
the limit $R \to \infty$ is reached with sufficient accuracy at 
finite screening radii $R$. 
The presence of the bound-state wave function in the matrix element
strongly suppresses the contribution of the screened Coulomb interaction 
in high partial waves, i.e., two-baryon partial waves with orbital 
angular momentum $L \le 6$ are sufficient for convergence.
The other quantum-number related cutoffs in the partial-wave dependence
of the matrix element are the same as in \Refs~\cite{deltuva:04a,deltuva:04b},
i.e.,  $I \le 4$, $\mathcal{J} \le \frac{15}{2}$ for photoreactions, 
and  $I \le 3$, $\mathcal{J} \le \frac{35}{2}$ for 
two-body electrodisintegration of $\He$. All calculations include
both total isospin $\mathcal{T} = \frac12$
and $\mathcal{T} = \frac32$ states.

\begin{figure}[!]
\begin{center}
\includegraphics[scale=\scl]{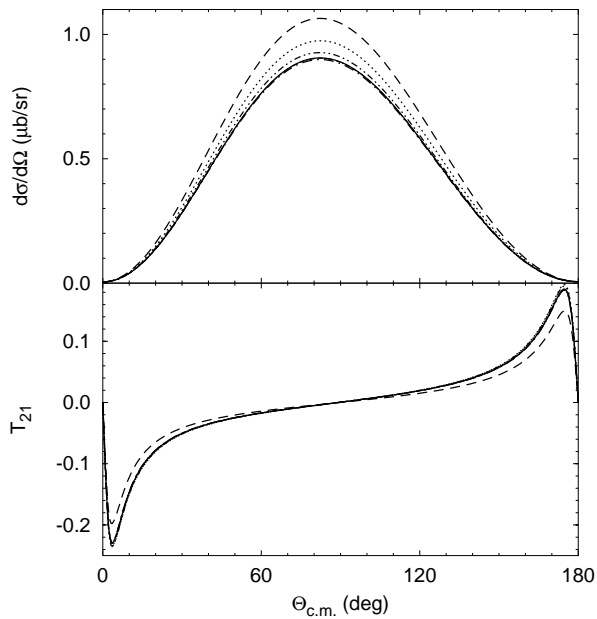}
\end{center}
\caption{\label{fig:rc3R}
Convergence of the differential cross section and of the deuteron 
analyzing power $T_{21}$ for $pd$ 
radiative capture at 3~MeV proton lab energy with screening radius $R$.
The observables are shown as functions of the c.m. scattering angle. 
The curves are explained in the caption of \Fig~\ref{fig:pd3R}.}
\end{figure}

Figure~\ref{fig:rc3R} studies the convergence 
of our method with increasing screening radius $R$
for $pd$ radiative capture at 3~MeV proton lab energy. 
We show the differential cross section and the deuteron tensor analyzing power 
$T_{21}$ which are the most critical observables in terms of 
convergence behavior.
As in the case of $pp$ and $pd$ elastic scattering the convergence 
is impressive and becomes more rapid with increasing energy;
it is quite comparable to $pd$ elastic scattering.
The convergence with increasing screening radius $R$ is the same for 
two-body electrodisintegration of $\He$; we therefore omit a
corresponding figure.

\subsection{Conclusions on the practical implementation of the 
Coulomb interaction}

Using the described method we are able to include the Coulomb interaction
between two protons in the description of hadronic and e.m. 
three-nucleon reactions in the $pd$ c.m. energy regime from about 1 MeV 
up to the pion production threshold. The screening radius 
required for the convergence decreases with increasing energy.
Whereas $R=20\fm$ is required for energies around deuteron breakup
threshold, the screening radius can be lowered to $R=10\fm$
above 10 MeV c.m. energy. In contrast, the screening radius has to be
increased considerably when calculating 
extreme low-energy quantities, such as the $pd$ doublet scattering length,
which at present is outside the reach of our adopted technique.
On the other hand, the high-energy limit is imposed by the form of the 
hadronic interaction which is applicable only below pion production threshold.
We notice no particular
feature of the convergence when crossing the three-body breakup threshold.
However, the paper does not treat the three-body breakup reactions yet.

\section{Results \label{sec:res}}

We base our calculations on the two-baryon coupled-channel potential
CD Bonn + $\Delta$ with and without Coulomb and use the CD Bonn potential
with and without Coulomb as purely nucleonic reference. We use the charge 
and current operators of \Refs~\cite{deltuva:04a,deltuva:04b},
appropriate for the underlying dynamics. We add relativistic corrections
to the charge in the Siegert part of the operator when describing 
the photoreactions, an admittedly questionable procedure, 
but entirely unrelated to the real issue of Coulomb in this paper.

Obviously, we have much more predictions than it is possible and wise
to show. Therefore we make a judicious selection and present those
predictions which illustrate the message we believe the results tell us.
The readers, dissatisfied with our choice, are welcome to obtain the results
for their favorite data from us.

Our predictions are dominantly based on the two-baryon coupled-channel 
potential CD Bonn + $\Delta$; its single virtual $\Delta$-isobar excitation
yields, in the three-nucleon system, an effective three-nucleon force
consistent with the two-nucleon interaction. 
$\Delta$-isobar effects increase the $\Hh$ binding energy from
8.004 MeV for CD Bonn to 8.297 MeV for CD Bonn + $\Delta$,
the experimental value being 8.482 MeV. That binding energy increase has
simultaneous beneficial effects on other bound state properties, e.g.,
on the charge radius, but those effects also appear in the $pd$ elastic
scattering amplitude, especially in the three-nucleon
$\mathcal{J}^\Pi = \frac12 ^+$ partial wave. The correlation between
trinucleon binding and other low-energy observables is known as scaling.
However, the Coulomb interaction also makes a significant contribution
to trinucleon binding; the $\Hh - \He$ 
binding energy difference is 0.746 MeV for CD Bonn and 0.756 MeV for
CD Bonn + $\Delta$, compared with the experimentally required value
of 0.764 MeV. This binding energy difference is dominantly due to the 
Coulomb repulsion between the protons in $\He$; the contribution arising
from the hadronic charge asymmetry is much smaller.

In three-nucleon scattering the $\Delta$ isobar therefore contributes
to the scaling phenomenon at low energies, but it manifests itself more
directly at higher energies when channel coupling becomes more probable.
This section tries to explore the interplay between $\Delta$-isobar
and Coulomb effects in the considered three-nucleon reactions.

\subsection{Elastic proton-deuteron scattering}

Figures~\ref{fig:pd3} and \ref{fig:pd9} give characteristic low-energy 
results.  As examples we show observables
at 3~MeV and 9~MeV proton lab energy, respectively below and above
deuteron breakup threshold. 
The Coulomb effect is quite appreciable at all scattering angles, 
but its relative importance decreases with increasing energy.
In contrast, on the scale of the observed Coulomb effect, the 
$\Delta$-isobar effect is minute  at those low energies.
The inclusion of  Coulomb is essential for a successful  account of data
for the spin-averaged differential cross section and for the deuteron
tensor analyzing powers. 
However, the inclusion of Coulomb increases the discrepancy between
theoretical predictions and experimental data
in the peak region of proton and deuteron vector analyzing powers,
the so-called $A_y$-puzzle. 
Our findings are consistent with the results of 
\Refs~\cite{alt:02a,kievsky:01a}.

Figure~\ref{fig:pd135} shows selected results at 135 MeV proton lab energy.
The Coulomb effect is confined to the forward direction, i.e.,
to c.m. scattering angles smaller than 30 degrees
where the $\Delta$-isobar effect is not visible. 
The $\Delta$-isobar effect shows up rather strongly in the region of the
diffraction minimum, where its effect is beneficial and the Coulomb effect
is gone. $\Delta$-isobar and Coulomb effects are nicely separated.
Thus, the $\Delta$-isobar effect found previously \cite{deltuva:03c}
on the Sagara discrepancy and on spin 
observables remains essentially unchanged by the 
inclusion of the Coulomb interaction.
The predictions of \Fig~\ref{fig:pd135} are characteristic for all observables
at higher energies.

\renewcommand{\scl}{0.5}
\begin{figure*}[!]
\begin{center}
\includegraphics[scale=\scl]{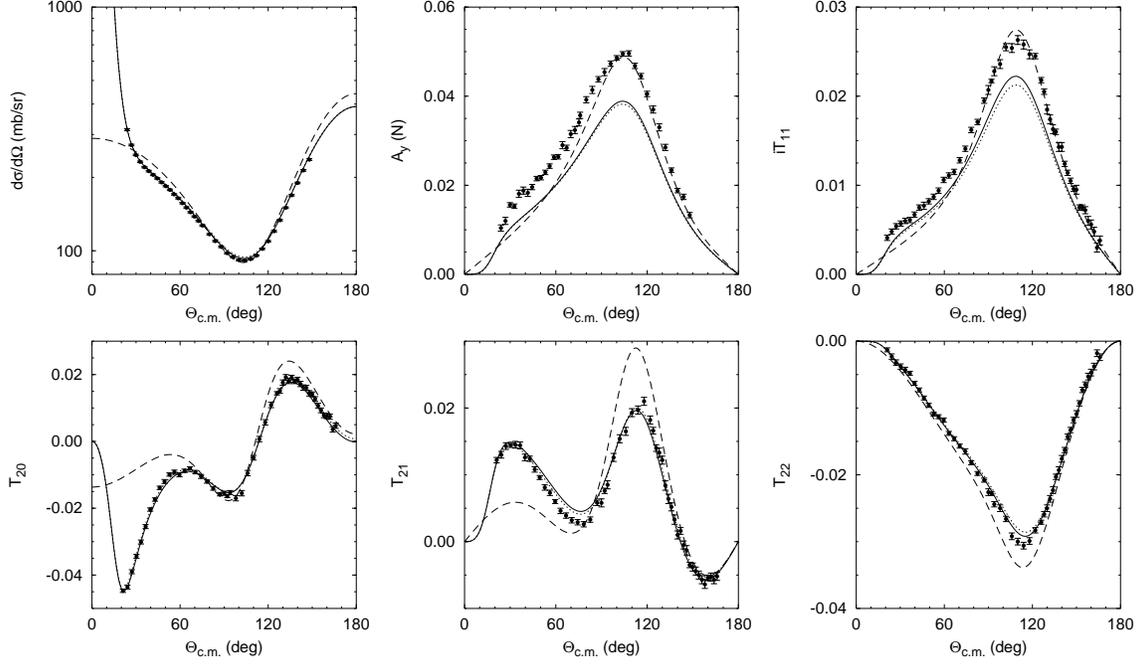}
\end{center}
\caption{\label{fig:pd3}
Differential cross section and analyzing powers for $pd$ elastic scattering 
at 3~MeV proton lab energy as functions of the c.m. scattering angle. 
Results including $\Delta$-isobar excitation and the Coulomb interaction
(solid curves) are compared to results without Coulomb (dashed curves).
In order to appreciate the size of the $\Delta$-isobar effect the purely
nucleonic results including Coulomb are also shown (dotted curves).
The experimental data are from \Ref~\cite{shimizu:95a}. } 
\end{figure*}

\begin{figure*}[!]
\begin{center}
\includegraphics[scale=\scl]{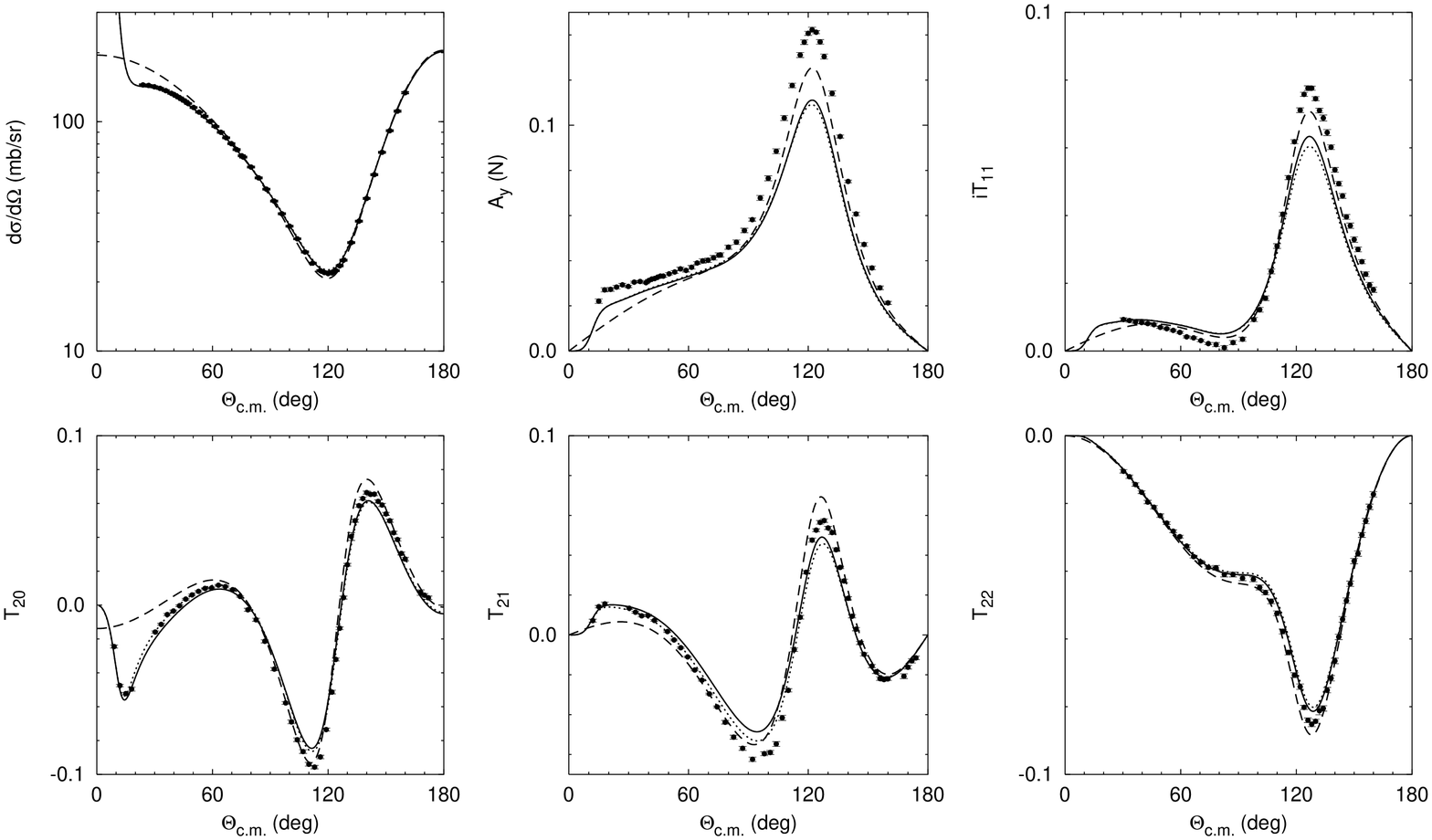}
\end{center}
\caption{\label{fig:pd9}
Differential cross section and analyzing powers for $pd$ elastic scattering 
at  9~MeV proton lab energy as functions of the c.m. scattering angle.
The curves are explained in the caption of \Fig~\ref{fig:pd3}.
The experimental data are from \Ref~\cite{sagara:94a}.} 
\end{figure*}

\begin{figure}[!]
\begin{center}
\includegraphics[scale=0.63]{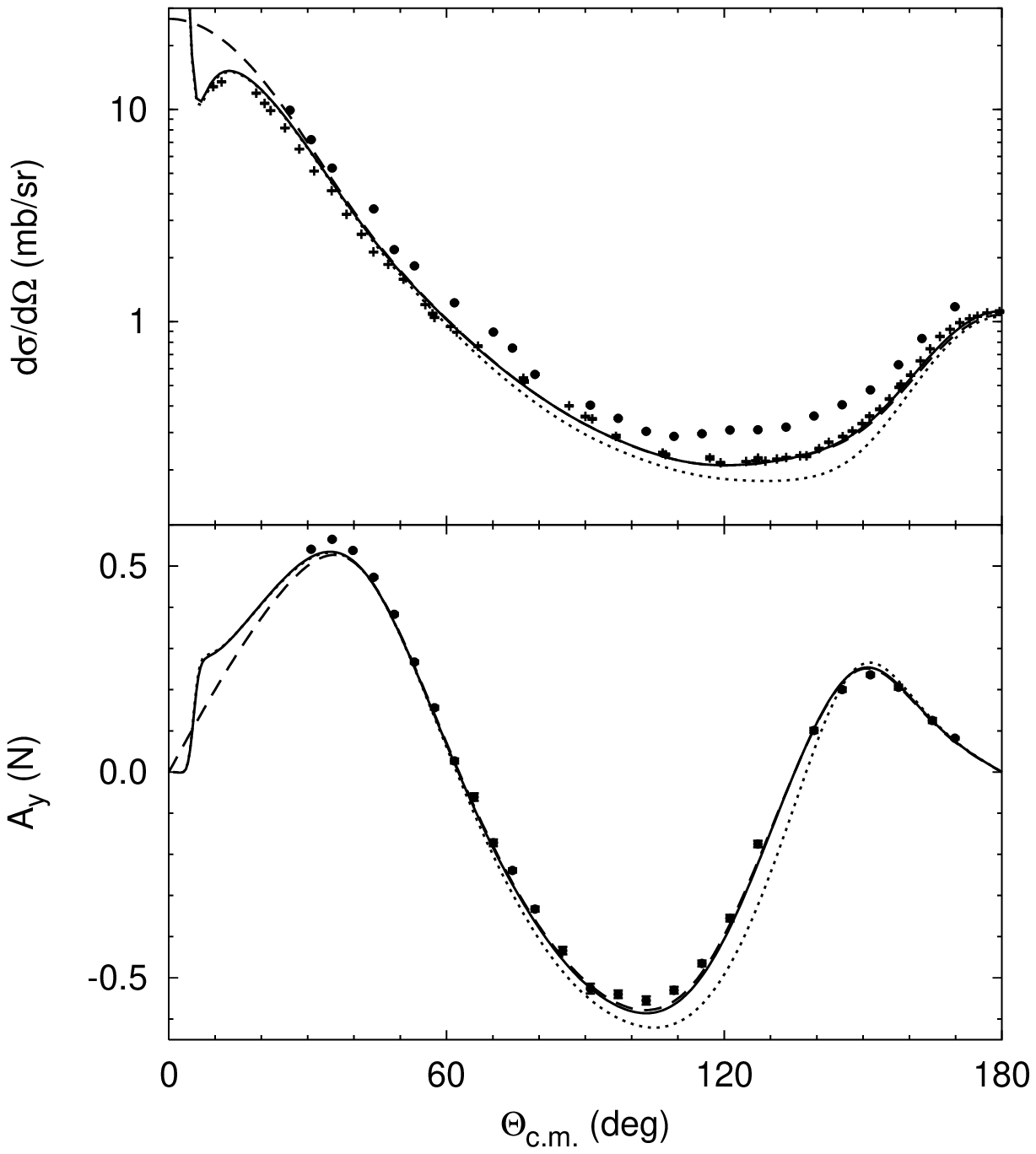}
\end{center}
\caption{\label{fig:pd135}
Differential cross section and proton analyzing power $A_y(N)$
for $pd$ elastic scattering 
at 135~MeV proton lab energy as functions of the c.m. scattering angle.
The curves are explained in the caption of \Fig~\ref{fig:pd3}.
The experimental data  are from 
\Ref~\cite{sekiguchi:02a} (crosses) and from \Ref~\cite{ermisch:03b}
(full circles) for the differential cross section,
and from \Ref~\cite{ermisch:01a} for the analyzing power.}
\end{figure}

\subsection{Proton-deuteron radiative capture}

The e.m. current operator is the standard choice of \Ref~\cite{deltuva:04a}
supplemented  by the relativistic one-nucleon charge
corrections, also given in \Ref~\cite{deltuva:04a},
 which we found to be important for some spin observables
even at low energies.

References \cite{marcucci:05a,golak:00a} carried out corresponding
realistic calculations for $pd$ radiative capture with 
different two-nucleon potentials and an irreducible three-nucleon force,
but without relativistic one-nucleon charge corrections. The
calculations of \Ref~\cite{marcucci:05a} take the Coulomb interaction fully
into account, but are limited to reactions below 10 MeV c.m. energy.
Reference \cite{golak:00a} neglects the Coulomb interaction
in the continuum states. 
When comparable, the results of this paper and those of 
\Refs~\cite{marcucci:05a,golak:00a} agree qualitatively;
benchmark comparisons have not been done.

\begin{figure*}[!] 
\begin{center}
\includegraphics[scale=\scl]{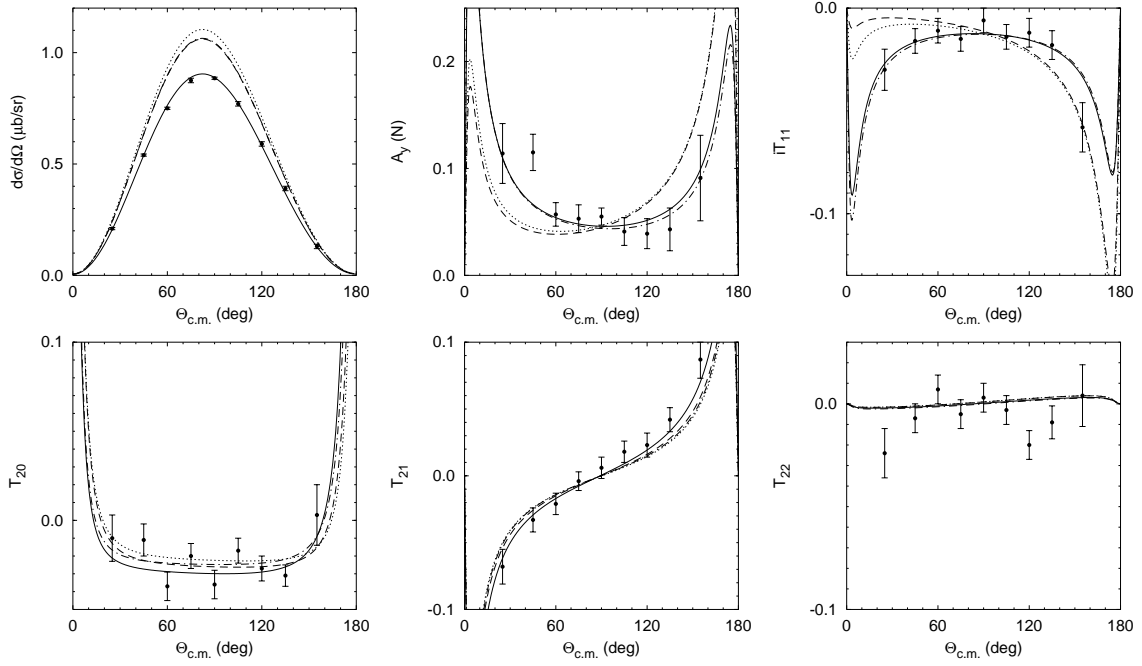}
\end{center}
\caption{\label{fig:rc3}
Differential cross section and analyzing powers for $pd$ radiative capture 
at 3~MeV proton lab energy as functions of the photon c.m. scattering angle
with respect to the direction of the proton. 
The shown results include, respectively: 
Coulomb interaction, relativistic one-nucleon charge correction, and
 $\Delta$-isobar excitation (solid curves);  
relativistic one-nucleon charge corrections and $\Delta$-isobar excitation 
(dashed-dotted curves);
$\Delta$-isobar excitation (dashed curves);
purely nucleonic results (dotted curves).
The experimental data are from \Ref~\cite{smith:99a}.} 
\end{figure*}

Figure~\ref{fig:rc3} shows the Coulomb effect for $pd$ radiative
capture at 3~MeV proton lab energy. The Coulomb effect is most
important for the differential cross section which is reduced by about
20\% and agrees rather well with the experimental data. 
In contrast, the spin observables show only a small Coulomb effect.
The effect of relativistic one-nucleon charge corrections
is entirely negligible for the differential cross section, but
rather sizable  and necessary for a satisfactory description of the data
for the vector analyzing powers. Our results without 
relativistic one-nucleon charge corrections are consistent with the
corresponding calculations of \Ref~\cite{marcucci:05a} which also
fail to account for the vector analyzing power data.
A moderate $\Delta$-isobar effect due to scaling is visible around the 
peak of the differential cross section.

\begin{figure*}[!]
\begin{center}
\includegraphics[scale=\scl]{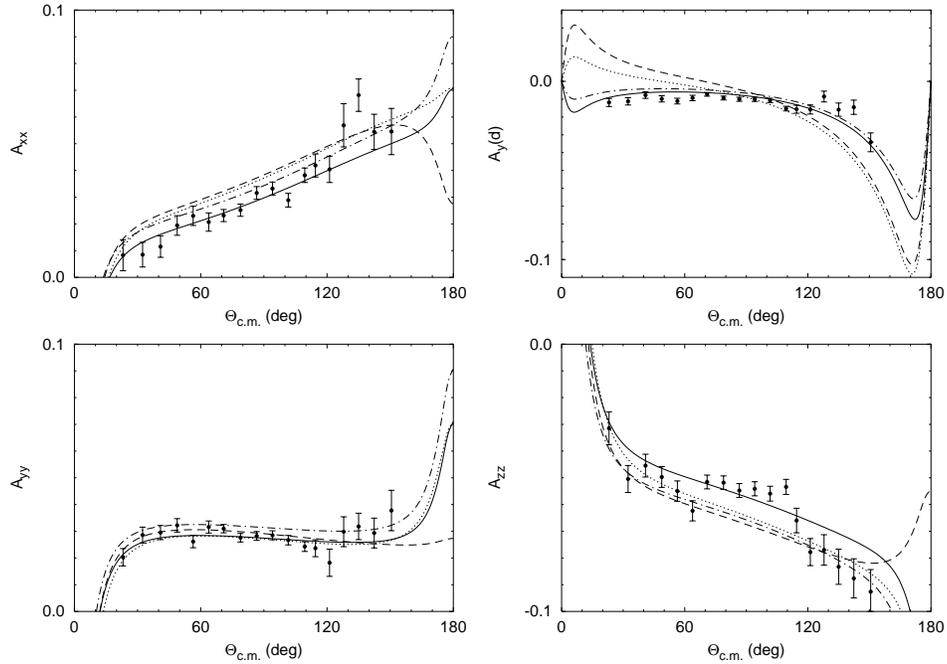}
\end{center}
\caption{\label{fig:rc17.5d}
Deuteron analyzing powers for $pd$ radiative capture at 17.5~MeV deuteron 
lab energy as functions of the photon c.m. scattering angle
with respect to the direction of the proton.
The curves are explained in the caption of \Fig~\ref{fig:rc3}.
The experimental data are from \Ref~\cite{akiyoshi:01a}. } 
\end{figure*}

Selected deuteron analyzing powers at 17.5~MeV 
deuteron lab energy with  moderate Coulomb effects are shown 
in \Fig~\ref{fig:rc17.5d} together with the experimental data. 
Since the deuteron analyzing power $A_{yy}$ is rather flat
between 40~deg and 140~deg according to \Fig~\ref{fig:rc17.5d}, 
\Fig~\ref{fig:Ayy90} focuses on the energy dependence
of  $A_{yy}$  at 90~deg photon lab scattering  angle.
Clearly,  our calculation accounts rather well for the known data of $A_{yy}$
in the whole deuteron lab energy region  up to 95 MeV. However,
a similar study for $T_{20}$ at 90~deg (not shown here)
indicates that the strong energy dependence of the low energy data from 
TUNL \cite{schmid:96a} is not compatible with the present calculation.
The rather good agreement with experimental data in \Figs~\ref{fig:rc17.5d} 
and \ref{fig:Ayy90} is obtained in general as an interplay of three
considered effects, i.e., the effects due to the $\Delta$ isobar, 
due to the relativistic one-nucleon charge corrections, and due to
the Coulomb interaction.

Figure~\ref{fig:rc150} shows the differential cross section
and the nucleon analyzing power for $pd$ radiative capture at
150~MeV nucleon lab energy where we previously found  rather
significant $\Delta$-isobar effects \cite{deltuva:04a}.
Even at this relatively high energy there is a visible, though small
Coulomb effect around the peak of the differential cross section.
In addition, both observables show a sizable effect of the relativistic 
one-nucleon charge corrections.

Figures \ref{fig:rc3} - \ref{fig:rc150}
also recall the non-Coulomb effects on observables in order to discuss 
their interplay with Coulomb. In the differential cross section
at 3 MeV and 150 MeV proton lab energy the $\Delta$ isobar plays
different roles; at low energy the $\Delta$ isobar manifests itself
through scaling due to changed bound state properties,
but at higher energies the explicit excitation of $\Delta$ channels 
in scattering becomes more predominant.
We also note that relativistic one-nucleon charge corrections are important
for proton and deuteron vector analyzing powers even at low energies;
with increasing energy they become quite significant in general.
The relativistic one-nucleon charge corrections are mostly beneficial 
for the account of the experimental data,
though their inclusion is not fully consistent with the 
underlying nonrelativistic hadronic dynamics.

\renewcommand{\scl}{0.63}
\begin{figure}[!]
\begin{center}
\includegraphics[scale=\scl]{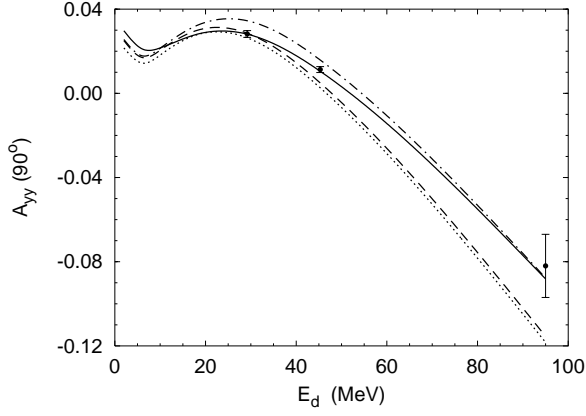}
\end{center}
\caption{\label{fig:Ayy90}
Deuteron analyzing power $A_{yy}$ for $pd$ radiative capture at 90~deg photon
lab scattering  angle as functions of the deuteron lab energy.
The curves are explained in the caption of \Fig~\ref{fig:rc3}.
The experimental data are from \Refs~\cite{jourdan:85a,pitts:88a}. } 
\end{figure}

\begin{figure}[!]
\begin{center}
\includegraphics[scale=\scl]{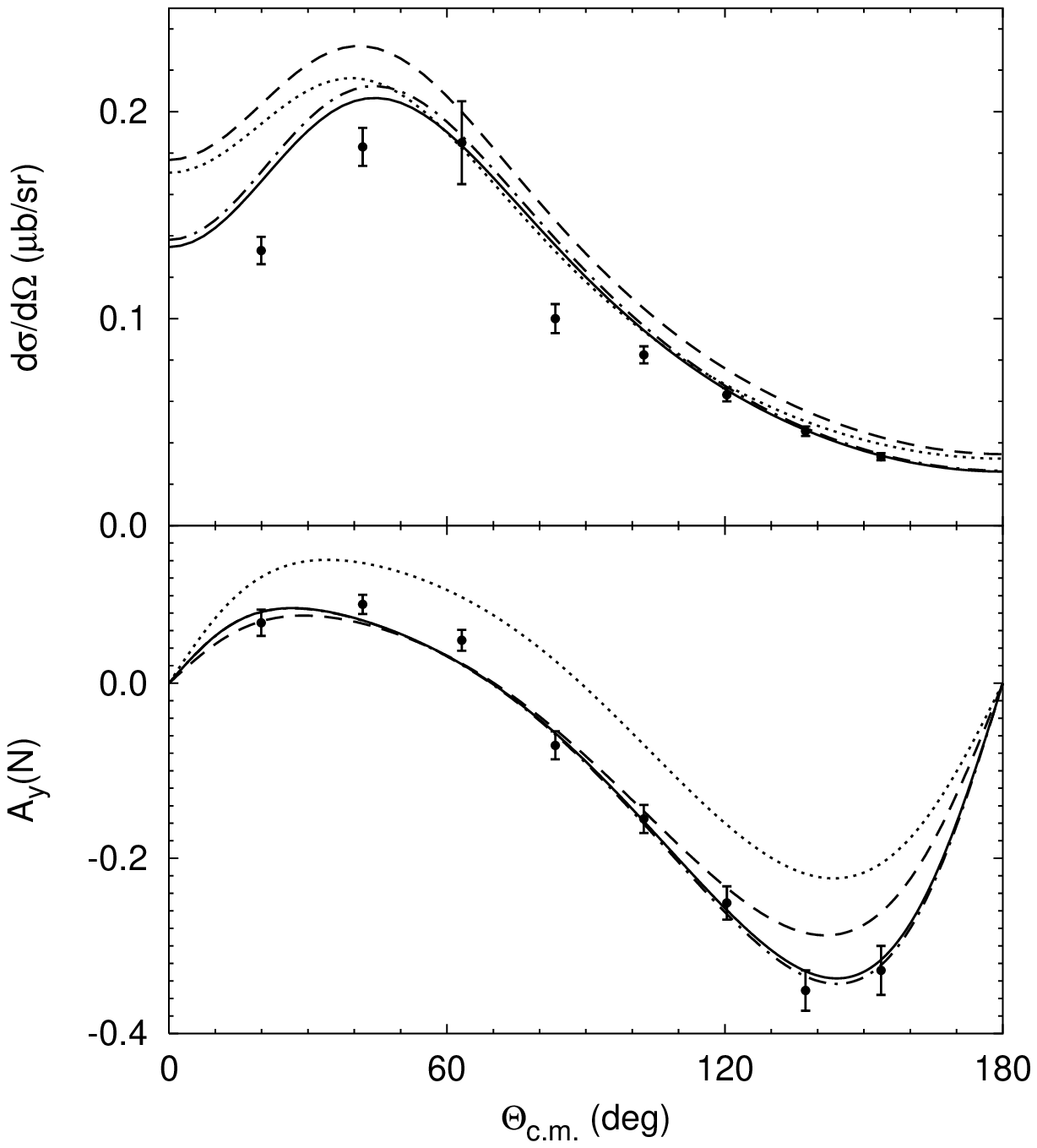}
\end{center}
\caption{\label{fig:rc150}
Differential cross section and proton analyzing power $A_{y}(N)$ 
for $pd$ radiative capture at 150~MeV proton lab energy 
as functions of the photon c.m. scattering angle
with respect to the direction of the proton. 
The curves are explained in the caption of \Fig~\ref{fig:rc3}.
The experimental data are from \Ref~\cite{pickar:87a}. } 
\end{figure}

\subsection{Two-body electrodisintegration of $\He$}

\begin{figure}[!]
\begin{center}
\includegraphics[scale=\scl]{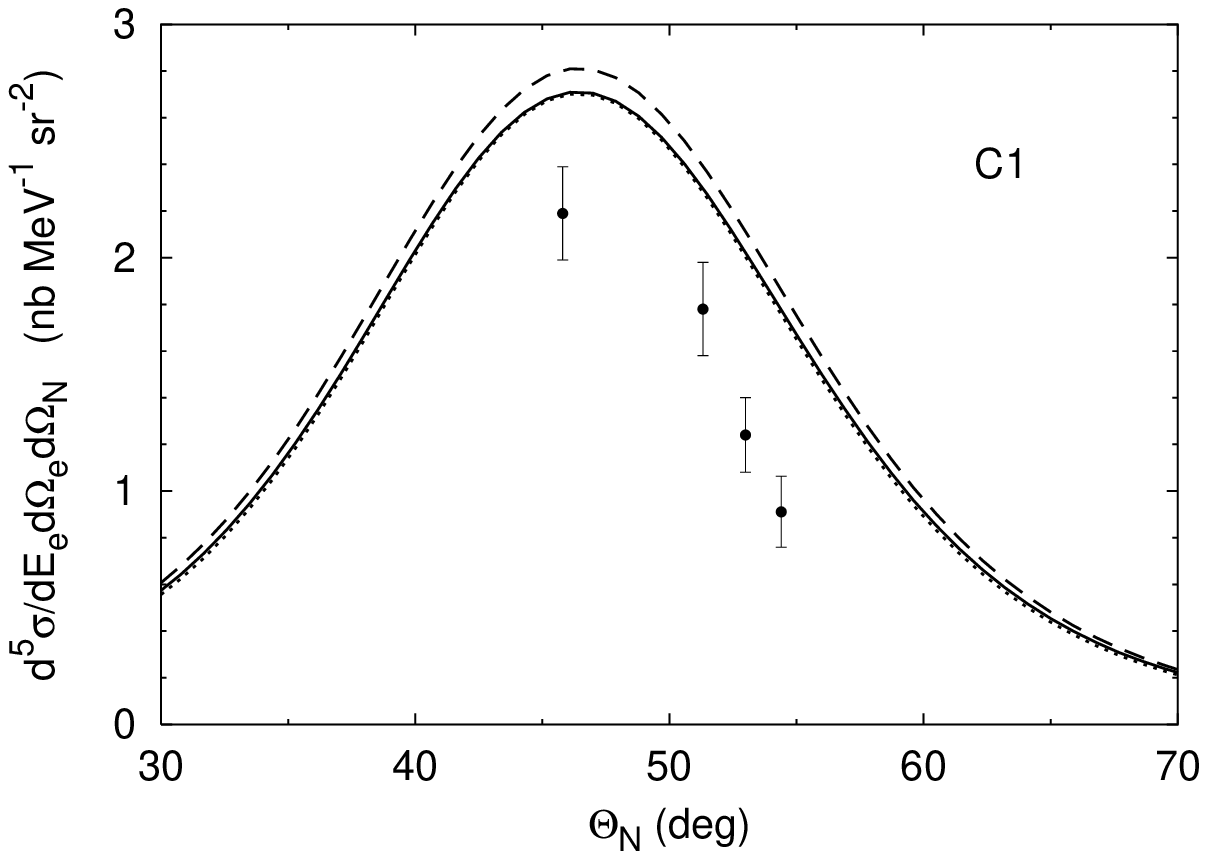} \\
\includegraphics[scale=\scl]{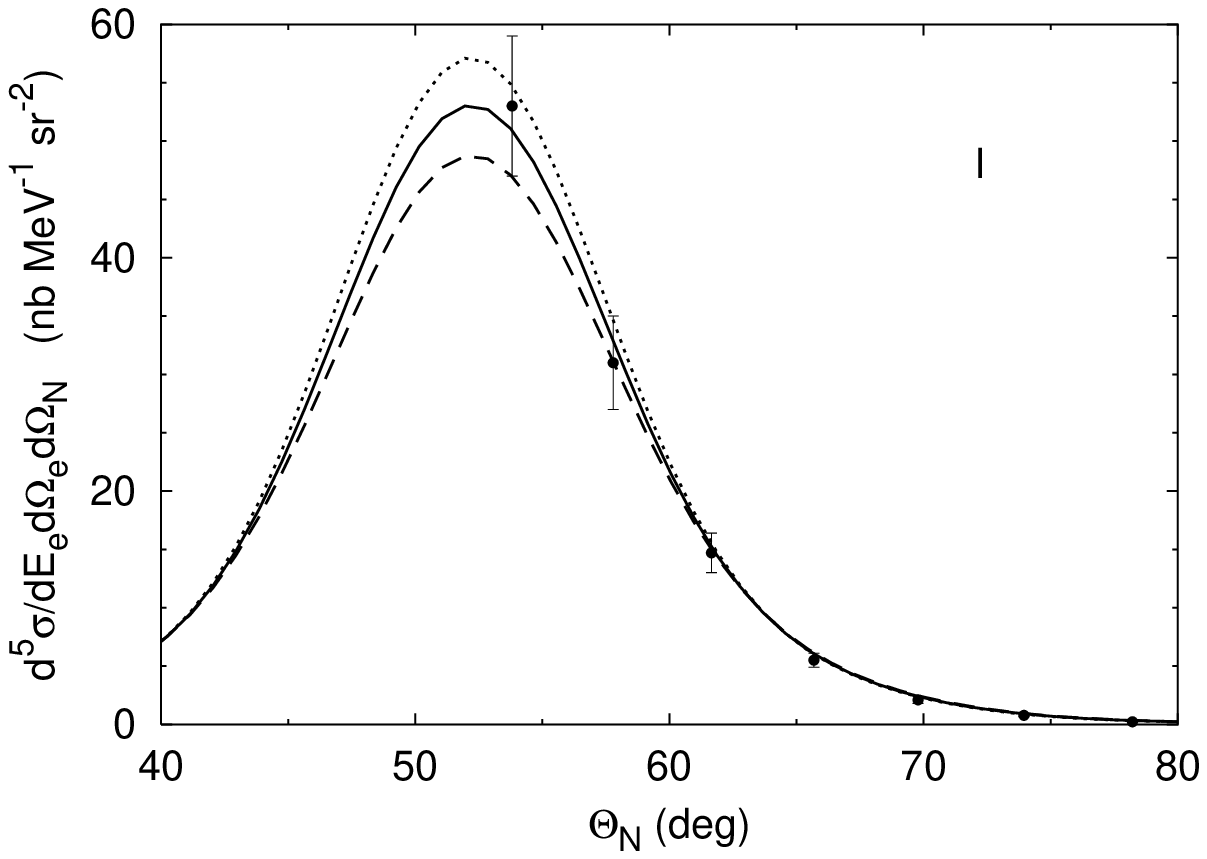} \\
\includegraphics[scale=\scl]{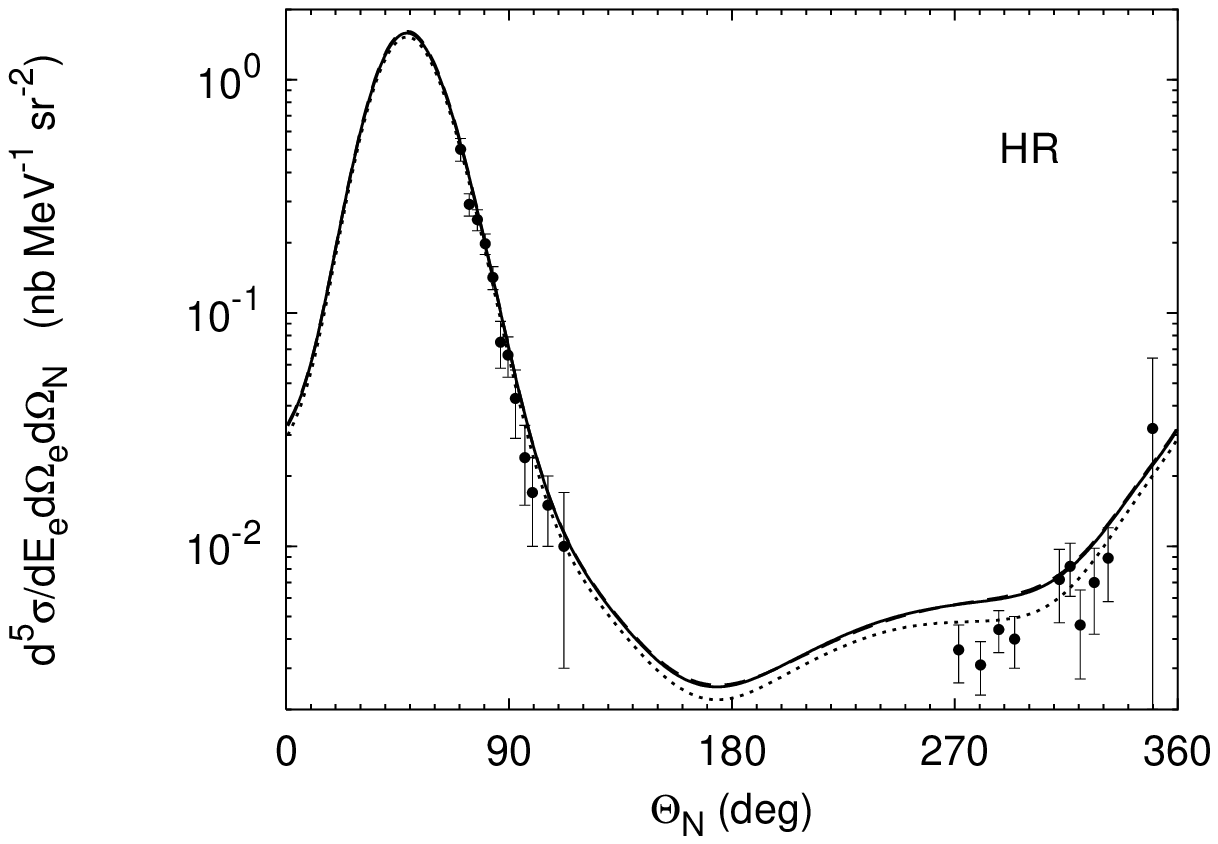} 
\end{center}
\caption{\label{fig:eB}
Lab differential cross section for two-body electrodisintegration of $\He$
as function of the proton lab scattering angle.
 The electron lab energy, scattering angle, the momentum and energy transfer 
are 390 MeV, 74.4 deg, 434.8 MeV and 66.1 MeV for the reaction kinematics C1,
527.9 MeV, 52.2 deg, 430.0 MeV and 99.8 MeV for the reaction kinematics I,
and 390 MeV, 39.7 deg, 250.2 MeV and 113.0 MeV for the reaction kinematics HR
of \Ref~\cite{jans:87a}, respectively.
Results including $\Delta$-isobar excitation and the Coulomb interaction
(solid curves) are compared to the results without Coulomb (dashed curves).
In order to appreciate the size of the $\Delta$-isobar effect the purely
nucleonic results with Coulomb are also shown (dotted curves).
The experimental data are from \Ref~\cite{jans:87a}.} 
\end{figure}

The e.m. current operator is taken over from \Ref~\cite{deltuva:04b};
compared to photo reactions, the relativistic one-nucleon charge
corrections are less important and are therefore omitted.
The Coulomb effect in the two-body electrodisintegration of $\He$
depends on both energy and momentum transfer. We do not study
that dependence in detail. We only show in
\Fig~\ref{fig:eB} a sample result for the three reaction
kinematics C1, I and HR of \Ref~\cite{jans:87a}.
The Coulomb effect on the C1 and I differential cross sections
is visible in the peak,  though in C1 small compared with the 
discrepancy between theoretical predictions and experimental data. 
In the HR differential cross section a Coulomb effect is not
visible in the logarithmic scale of the plot, but instead a
$\Delta$-isobar effect at backward angles.
Qualitatively our results without Coulomb
agree well with the ones of \Ref~\cite{ishikawa:94a}.

\section{Summary \label{sec:concl}}

This paper shows how the Coulomb interaction between the charged baryons
can be included into the momentum-space description of elastic 
proton-deuteron scattering and of related e.m. reactions 
using the screening and renormalization approach.
The theoretical framework is the AGS integral equation \cite{alt:67a}.
The calculations are done on the same level of accuracy and 
sophistication as for the corresponding neutron-deuteron reactions.
The conclusions of the paper refer to the developed technique and
to the physics results obtained with that technique.

\emph{Technically}, the idea of screening and renormalization is the one of
\Refs~\cite{taylor:74a,alt:78a} and 
we rely on these works for mathematical rigor.
However, our practical realization  differs quite significantly
from the one of \Refs~\cite{berthold:90a,alt:02a}:

(1) We use modern hadronic interactions, CD Bonn and CD Bonn + $\Delta$, 
in contrast to the low-rank separable potentials
of \Refs~\cite{berthold:90a,alt:02a}. Our use of the full  potential
requires the standard form of the three-particle equations, 
different from the quasiparticle approach of \Refs~\cite{berthold:90a,alt:02a}.

(2) We do not approximate the screened Coulomb transition 
matrix by the screened Coulomb potential. 

(3) The quasiparticle approach of \Refs~\cite{berthold:90a,alt:02a} treats the
screened Coulomb potential between the protons without partial-wave 
expansion and therefore has no problems with the slow convergence
of that expansion. Our solution of three-nucleon equations proceeds
in partial-wave basis and therefore is faced with the slow partial-wave
convergence of the Coulomb interaction between the charged baryons.
However, we are able to
obtain fully converged results by choosing a special form of the screening
function and by using the perturbation theory of \Ref~\cite{deltuva:03b} 
for treating the screened Coulomb transition matrix in high partial waves.
This would not be possible if we had used Yukawa screening as in 
\Refs~\cite{berthold:90a,alt:02a}.

(4) Our method for including the Coulomb interaction is efficient.
Though the number of the isospin triplet partial waves to be taken into account
is considerably higher than in the case without Coulomb,
the required computing time increases only by a factor of 2 to 3,
due to the use of the perturbation theory for high partial waves.

The obtained results are stable and well checked for their validity.
The employed technique gets cumbersome when approaching very low energies,
i.e., $pd$ c.m. energies below 1~MeV, due to the need for very large 
screening radii. Thus, the calculation of the
doublet scattering length for elastic $pd$ scattering is, at present,
outside of our numerical reach, a barrier which does not exist for the
coordinate-space techniques adopted in 
\Refs~\cite{kievsky:01a,chen:01a}.
On the other hand, we do not see any particular numerical problem
when crossing the breakup threshold and going to higher energies
where coordinate-space techniques are very hard to apply.

The present technique is not yet used for breakup itself, but such an
 extension is on its way.

\emph{Physicswise}, the Coulomb effect in elastic $pd$ scattering
is important at low energies for all
kinematic regimes, but gets confined to the forward direction at 
higher energies, whereas the  effect mediated by the $\Delta$ isobar 
remains almost unmodified by the inclusion of Coulomb.
In radiative $pd$ capture the Coulomb effect is important for low-energy
differential cross sections and for some spin observables up to about
30~MeV proton lab energy; at higher energies there is still a
visible Coulomb effect for some observables, e.g., 
in the peak of the differential cross section. 
In two-body electrodisintegration of $\He$ 
the Coulomb effect appears not to be simply related to the internal
excitation of the three-nucleon system. 
A thorough study of the dependence of the Coulomb effect on
the energy- and three-momentum transfer to the $\He$ 
target is beyond the scope of this paper.

\begin{acknowledgments}
The authors thank A.~Kievsky and R.~Lazauskas for providing
benchmark results, K.~Sagara for providing the experimental data,
and S.~Oryu and A.~Stadler for useful discussions.
A.D. is supported by the FCT grant SFRH/BPD/14801/2003 
and by the DFG grant Sa 247/25, P.U.S. in part by the DFG grant Sa 247/25,
and A.C.F. in part by the grant POCTI/FNU/37280/2001. 
\end{acknowledgments}

%%%%%%%%%%%%%%%%%%%%%%%%%%%%%%%%%%%%%%%%%%%%%%%%%%%%%%%%%%%%%%%%%%%%%%%%%%%%%
%\clearpage

%\begin{appendix}\end{appendix}
%%%%%%%%%%%%%%%%%%%%%%%%%%%%%%%%%%%%%%%%%%%%%%%%%%%%%%%%%%%%%%%%%%%%%%%%%%%%%
%\bibliographystyle{prsty}
%\bibliography{abbrev,hann,book,pre80,80-89,90-99,200x,publication,exp,clmb}

\begin{thebibliography}{36}
\expandafter\ifx\csname natexlab\endcsname\relax\def\natexlab#1{#1}\fi
\expandafter\ifx\csname bibnamefont\endcsname\relax
  \def\bibnamefont#1{#1}\fi
\expandafter\ifx\csname bibfnamefont\endcsname\relax
  \def\bibfnamefont#1{#1}\fi
\expandafter\ifx\csname citenamefont\endcsname\relax
  \def\citenamefont#1{#1}\fi
\expandafter\ifx\csname url\endcsname\relax
  \def\url#1{\texttt{#1}}\fi
\expandafter\ifx\csname urlprefix\endcsname\relax\def\urlprefix{URL }\fi
\providecommand{\bibinfo}[2]{#2}
\providecommand{\eprint}[2][]{\url{#2}}

\bibitem[{\citenamefont{Alt et~al.}(2002)\citenamefont{Alt, Mukhamedzhanov,
  Nishonov, and Sattarov}}]{alt:02a}
\bibinfo{author}{\bibfnamefont{E.~O.} \bibnamefont{Alt}},
  \bibinfo{author}{\bibfnamefont{A.~M.} \bibnamefont{Mukhamedzhanov}},
  \bibinfo{author}{\bibfnamefont{M.~M.} \bibnamefont{Nishonov}},
  \bibnamefont{and} \bibinfo{author}{\bibfnamefont{A.~I.}
  \bibnamefont{Sattarov}}, \bibinfo{journal}{Phys.~Rev.~C}
  \textbf{\bibinfo{volume}{65}}, \bibinfo{pages}{064613}
  (\bibinfo{year}{2002}).

\bibitem[{\citenamefont{Berthold et~al.}(1990)\citenamefont{Berthold, Stadler,
  and Zankel}}]{berthold:90a}
\bibinfo{author}{\bibfnamefont{G.~H.} \bibnamefont{Berthold}},
  \bibinfo{author}{\bibfnamefont{A.}~\bibnamefont{Stadler}}, \bibnamefont{and}
  \bibinfo{author}{\bibfnamefont{H.}~\bibnamefont{Zankel}},
  \bibinfo{journal}{Phys.~Rev.~C} \textbf{\bibinfo{volume}{41}},
  \bibinfo{pages}{1365} (\bibinfo{year}{1990}).

\bibitem[{\citenamefont{Kievsky et~al.}(2001)\citenamefont{Kievsky, Viviani,
  and Rosati}}]{kievsky:01a}
\bibinfo{author}{\bibfnamefont{A.}~\bibnamefont{Kievsky}},
  \bibinfo{author}{\bibfnamefont{M.}~\bibnamefont{Viviani}}, \bibnamefont{and}
  \bibinfo{author}{\bibfnamefont{S.}~\bibnamefont{Rosati}},
  \bibinfo{journal}{Phys.~Rev.~C} \textbf{\bibinfo{volume}{64}},
  \bibinfo{pages}{024002} (\bibinfo{year}{2001}).

\bibitem[{\citenamefont{Chen et~al.}(2001)\citenamefont{Chen, Friar, and
  Payne}}]{chen:01a}
\bibinfo{author}{\bibfnamefont{C.~R.} \bibnamefont{Chen}},
  \bibinfo{author}{\bibfnamefont{J.~L.} \bibnamefont{Friar}}, \bibnamefont{and}
  \bibinfo{author}{\bibfnamefont{G.~L.} \bibnamefont{Payne}},
  \bibinfo{journal}{Few-Body Syst.} \textbf{\bibinfo{volume}{31}},
  \bibinfo{pages}{13} (\bibinfo{year}{2001}).

\bibitem[{\citenamefont{Suslov and Vlahovic}(2004)}]{suslov:04a}
\bibinfo{author}{\bibfnamefont{V.~M.} \bibnamefont{Suslov}} \bibnamefont{and}
  \bibinfo{author}{\bibfnamefont{B.}~\bibnamefont{Vlahovic}},
  \bibinfo{journal}{Phys.~Rev.~C} \textbf{\bibinfo{volume}{69}},
  \bibinfo{pages}{044003} (\bibinfo{year}{2004}).

\bibitem[{\citenamefont{Alt et~al.}(2004)\citenamefont{Alt, Levin, and
  Yakovlev}}]{alt:04a}
\bibinfo{author}{\bibfnamefont{E.~O.} \bibnamefont{Alt}},
  \bibinfo{author}{\bibfnamefont{S.~B.} \bibnamefont{Levin}}, \bibnamefont{and}
  \bibinfo{author}{\bibfnamefont{S.~L.} \bibnamefont{Yakovlev}},
  \bibinfo{journal}{Phys.~Rev.~C} \textbf{\bibinfo{volume}{69}},
  \bibinfo{pages}{034002} (\bibinfo{year}{2004}).

\bibitem[{\citenamefont{Oryu}(2004)}]{oryu:04a}
\bibinfo{author}{\bibfnamefont{S.}~\bibnamefont{Oryu}},
  \bibinfo{journal}{Few-Body Syst.} \textbf{\bibinfo{volume}{34}},
  \bibinfo{pages}{113} (\bibinfo{year}{2004});
  S. Oryu and S. Gojuki, Prog.~Theor.~Phys.~Suppl. {\bf 154}, 285 (2004).

\bibitem[{\citenamefont{Gl\"ockle et~al.}(1996)\citenamefont{Gl\"ockle,
  Wita{\l}a, H\"uber, Kamada, and Golak}}]{gloeckle:96a}
\bibinfo{author}{\bibfnamefont{W.}~\bibnamefont{Gl\"ockle}},
  \bibinfo{author}{\bibfnamefont{H.}~\bibnamefont{Wita{\l}a}},
  \bibinfo{author}{\bibfnamefont{D.}~\bibnamefont{H\"uber}},
  \bibinfo{author}{\bibfnamefont{H.}~\bibnamefont{Kamada}}, \bibnamefont{and}
  \bibinfo{author}{\bibfnamefont{J.}~\bibnamefont{Golak}},
  \bibinfo{journal}{Phys.~Rep.} \textbf{\bibinfo{volume}{274}},
  \bibinfo{pages}{107} (\bibinfo{year}{1996}).

\bibitem[{\citenamefont{Deltuva
  et~al.}(2003{\natexlab{a}})\citenamefont{Deltuva, Chmielewski, and
  Sauer}}]{deltuva:03a}
\bibinfo{author}{\bibfnamefont{A.}~\bibnamefont{Deltuva}},
  \bibinfo{author}{\bibfnamefont{K.}~\bibnamefont{Chmielewski}},
  \bibnamefont{and} \bibinfo{author}{\bibfnamefont{P.~U.} \bibnamefont{Sauer}},
  \bibinfo{journal}{Phys.~Rev.~C} \textbf{\bibinfo{volume}{67}},
  \bibinfo{pages}{034001} (\bibinfo{year}{2003}{\natexlab{a}}).

\bibitem[{\citenamefont{Doleschall et~al.}(1982)\citenamefont{Doleschall,
  Gr\"uebler, Konig, Schmelzbach, Sperisen, and Jenny}}]{doleschall:82a}
\bibinfo{author}{\bibfnamefont{P.}~\bibnamefont{Doleschall}},
  \bibinfo{author}{\bibfnamefont{W.}~\bibnamefont{Gr\"uebler}},
  \bibinfo{author}{\bibfnamefont{V.}~\bibnamefont{Konig}},
  \bibinfo{author}{\bibfnamefont{P.~A.} \bibnamefont{Schmelzbach}},
  \bibinfo{author}{\bibfnamefont{F.}~\bibnamefont{Sperisen}}, \bibnamefont{and}
  \bibinfo{author}{\bibfnamefont{B.}~\bibnamefont{Jenny}},
  \bibinfo{journal}{Nucl.~Phys.~A} \textbf{\bibinfo{volume}{380}},
  \bibinfo{pages}{72} (\bibinfo{year}{1982}).

\bibitem[{\citenamefont{Taylor}(1974)}]{taylor:74a}
\bibinfo{author}{\bibfnamefont{J.~R.} \bibnamefont{Taylor}},
  \bibinfo{journal}{Nuovo Cimento} \textbf{\bibinfo{volume}{B23}},
  \bibinfo{pages}{313} (\bibinfo{year}{1974}); 
 M.~D. Semon and J.~R. Taylor, {\it ibid.} {\bf A26}, 48 (1975).

\bibitem[{\citenamefont{Alt et~al.}(1978)\citenamefont{Alt, Sandhas, and
  Ziegelmann}}]{alt:78a}
\bibinfo{author}{\bibfnamefont{E.~O.} \bibnamefont{Alt}},
  \bibinfo{author}{\bibfnamefont{W.}~\bibnamefont{Sandhas}}, \bibnamefont{and}
  \bibinfo{author}{\bibfnamefont{H.}~\bibnamefont{Ziegelmann}},
  \bibinfo{journal}{Phys.~Rev.~C} \textbf{\bibinfo{volume}{17}},
  \bibinfo{pages}{1981} (\bibinfo{year}{1978}); 
  E.~O. Alt and  W. Sandhas, {\it ibid.} {\bf 21}, 1733 (1980).

\bibitem[{\citenamefont{Machleidt}(2001)}]{machleidt:01a}
\bibinfo{author}{\bibfnamefont{R.}~\bibnamefont{Machleidt}},
  \bibinfo{journal}{Phys.~Rev.~C} \textbf{\bibinfo{volume}{63}},
  \bibinfo{pages}{024001} (\bibinfo{year}{2001}).

\bibitem[{\citenamefont{Deltuva
  et~al.}(2003{\natexlab{b}})\citenamefont{Deltuva, Machleidt, and
  Sauer}}]{deltuva:03c}
\bibinfo{author}{\bibfnamefont{A.}~\bibnamefont{Deltuva}},
  \bibinfo{author}{\bibfnamefont{R.}~\bibnamefont{Machleidt}},
  \bibnamefont{and} \bibinfo{author}{\bibfnamefont{P.~U.} \bibnamefont{Sauer}},
  \bibinfo{journal}{Phys.~Rev.~C} \textbf{\bibinfo{volume}{68}},
  \bibinfo{pages}{024005} (\bibinfo{year}{2003}{\natexlab{b}}).

\bibitem[{\citenamefont{Alt et~al.}(1967)\citenamefont{Alt, Grassberger, and
  Sandhas}}]{alt:67a}
\bibinfo{author}{\bibfnamefont{E.~O.} \bibnamefont{Alt}},
  \bibinfo{author}{\bibfnamefont{P.}~\bibnamefont{Grassberger}},
  \bibnamefont{and} \bibinfo{author}{\bibfnamefont{W.}~\bibnamefont{Sandhas}},
  \bibinfo{journal}{Nucl.~Phys.} \textbf{\bibinfo{volume}{B2}},
  \bibinfo{pages}{167} (\bibinfo{year}{1967}).

\bibitem[{\citenamefont{Rodberg and Thaler}(1967)}]{rodberg:67a}
\bibinfo{author}{\bibfnamefont{L.~S.} \bibnamefont{Rodberg}} \bibnamefont{and}
  \bibinfo{author}{\bibfnamefont{R.~M.} \bibnamefont{Thaler}},
  \emph{\bibinfo{title}{Introduction to the Quantum Theory of Scattering}}
  (\bibinfo{publisher}{Academic Press}, \bibinfo{address}{New York},
  \bibinfo{year}{1967}).

\bibitem[{\citenamefont{Deltuva
  et~al.}(2003{\natexlab{c}})\citenamefont{Deltuva, Chmielewski, and
  Sauer}}]{deltuva:03b}
\bibinfo{author}{\bibfnamefont{A.}~\bibnamefont{Deltuva}},
  \bibinfo{author}{\bibfnamefont{K.}~\bibnamefont{Chmielewski}},
  \bibnamefont{and} \bibinfo{author}{\bibfnamefont{P.~U.} \bibnamefont{Sauer}},
  \bibinfo{journal}{Phys.~Rev.~C} \textbf{\bibinfo{volume}{67}},
  \bibinfo{pages}{054004} (\bibinfo{year}{2003}{\natexlab{c}}).

\bibitem[{\citenamefont{Alt et~al.}(1998)\citenamefont{Alt, Mukhamedzhanov, and
  Sattarov}}]{alt:98a}
\bibinfo{author}{\bibfnamefont{E.~O.} \bibnamefont{Alt}},
  \bibinfo{author}{\bibfnamefont{A.~M.} \bibnamefont{Mukhamedzhanov}},
  \bibnamefont{and} \bibinfo{author}{\bibfnamefont{A.~I.}
  \bibnamefont{Sattarov}}, \bibinfo{journal}{Phys.~Rev.~Lett.}
  \textbf{\bibinfo{volume}{81}}, \bibinfo{pages}{4820} (\bibinfo{year}{1998}).

\bibitem[{\citenamefont{Deltuva et~al.}(2005)\citenamefont{Deltuva, Fonseca,
  Kievsky, Rosati, Sauer, and Viviani}}]{deltuva:05b}
\bibinfo{author}{\bibfnamefont{A.}~\bibnamefont{Deltuva}},
  \bibinfo{author}{\bibfnamefont{A.~C.} \bibnamefont{Fonseca}},
  \bibinfo{author}{\bibfnamefont{A.}~\bibnamefont{Kievsky}},
  \bibinfo{author}{\bibfnamefont{S.}~\bibnamefont{Rosati}},
  \bibinfo{author}{\bibfnamefont{P.~U.} \bibnamefont{Sauer}}, \bibnamefont{and}
  \bibinfo{author}{\bibfnamefont{M.}~\bibnamefont{Viviani}}
  (\bibinfo{year}{2005}), \bibinfo{note}{in preparation}.

\bibitem[{\citenamefont{Deltuva
  et~al.}(2004{\natexlab{a}})\citenamefont{Deltuva, Yuan, Adam~Jr., Fonseca,
  and Sauer}}]{deltuva:04a}
\bibinfo{author}{\bibfnamefont{A.}~\bibnamefont{Deltuva}},
  \bibinfo{author}{\bibfnamefont{L.~P.} \bibnamefont{Yuan}},
  \bibinfo{author}{\bibfnamefont{J.}~\bibnamefont{Adam~Jr.}},
  \bibinfo{author}{\bibfnamefont{A.~C.} \bibnamefont{Fonseca}},
  \bibnamefont{and} \bibinfo{author}{\bibfnamefont{P.~U.} \bibnamefont{Sauer}},
  \bibinfo{journal}{Phys.~Rev.~C} \textbf{\bibinfo{volume}{69}},
  \bibinfo{pages}{034004} (\bibinfo{year}{2004}{\natexlab{a}}).

\bibitem[{\citenamefont{Deltuva
  et~al.}(2004{\natexlab{b}})\citenamefont{Deltuva, Yuan, Adam~Jr., and
  Sauer}}]{deltuva:04b}
\bibinfo{author}{\bibfnamefont{A.}~\bibnamefont{Deltuva}},
  \bibinfo{author}{\bibfnamefont{L.~P.} \bibnamefont{Yuan}},
  \bibinfo{author}{\bibfnamefont{J.}~\bibnamefont{Adam~Jr.}}, \bibnamefont{and}
  \bibinfo{author}{\bibfnamefont{P.~U.} \bibnamefont{Sauer}},
  \bibinfo{journal}{Phys.~Rev.~C} \textbf{\bibinfo{volume}{70}},
  \bibinfo{pages}{034004} (\bibinfo{year}{2004}{\natexlab{b}}).

\bibitem[{\citenamefont{Shimizu et~al.}(1995)\citenamefont{Shimizu, Sagara,
  Nakamura, Maeda, Miwa, Nishimori, Ueno, Nakashima, and
  Morinobu}}]{shimizu:95a}
\bibinfo{author}{\bibfnamefont{S.}~\bibnamefont{Shimizu}},
  \bibinfo{author}{\bibfnamefont{K.}~\bibnamefont{Sagara}},
  \bibinfo{author}{\bibfnamefont{H.}~\bibnamefont{Nakamura}},
  \bibinfo{author}{\bibfnamefont{K.}~\bibnamefont{Maeda}},
  \bibinfo{author}{\bibfnamefont{T.}~\bibnamefont{Miwa}},
  \bibinfo{author}{\bibfnamefont{N.}~\bibnamefont{Nishimori}},
  \bibinfo{author}{\bibfnamefont{S.}~\bibnamefont{Ueno}},
  \bibinfo{author}{\bibfnamefont{T.}~\bibnamefont{Nakashima}},
  \bibnamefont{and} \bibinfo{author}{\bibfnamefont{S.}~\bibnamefont{Morinobu}},
  \bibinfo{journal}{Phys.~Rev.~C} \textbf{\bibinfo{volume}{52}},
  \bibinfo{pages}{1193} (\bibinfo{year}{1995}).

\bibitem[{\citenamefont{Sagara et~al.}(1994)\citenamefont{Sagara, Oguri,
  Shimizu, Maeda, Nakamura, Nakashima, and S.}}]{sagara:94a}
\bibinfo{author}{\bibfnamefont{K.}~\bibnamefont{Sagara}},
  \bibinfo{author}{\bibfnamefont{H.}~\bibnamefont{Oguri}},
  \bibinfo{author}{\bibfnamefont{S.}~\bibnamefont{Shimizu}},
  \bibinfo{author}{\bibfnamefont{K.}~\bibnamefont{Maeda}},
  \bibinfo{author}{\bibfnamefont{H.}~\bibnamefont{Nakamura}},
  \bibinfo{author}{\bibfnamefont{T.}~\bibnamefont{Nakashima}},
  \bibnamefont{and} \bibinfo{author}{\bibfnamefont{S.}~\bibnamefont{Morinobu}},
  \bibinfo{journal}{Phys.~Rev.~C} \textbf{\bibinfo{volume}{50}},
  \bibinfo{pages}{576} (\bibinfo{year}{1994}); 
  K.~Sagara (private communication).

\bibitem[{\citenamefont{Sekiguchi et~al.}(2002)\citenamefont{Sekiguchi, Sakai,
  Wita{\l}a, Gl\"ockle, Golak, Hatano, Kamada, Kato, Maeda, Nishikawa
  et~al.}}]{sekiguchi:02a}
\bibinfo{author}{\bibfnamefont{K.}~\bibnamefont{Sekiguchi}},
  \bibinfo{author}{\bibfnamefont{H.}~\bibnamefont{Sakai}},
  \bibinfo{author}{\bibfnamefont{H.}~\bibnamefont{Wita{\l}a}},
  \bibinfo{author}{\bibfnamefont{W.}~\bibnamefont{Gl\"ockle}},
  \bibinfo{author}{\bibfnamefont{J.}~\bibnamefont{Golak}},
  \bibinfo{author}{\bibfnamefont{M.}~\bibnamefont{Hatano}},
  \bibinfo{author}{\bibfnamefont{H.}~\bibnamefont{Kamada}},
  \bibinfo{author}{\bibfnamefont{H.}~\bibnamefont{Kato}},
  \bibinfo{author}{\bibfnamefont{Y.}~\bibnamefont{Maeda}},
  \bibinfo{author}{\bibfnamefont{J.}~\bibnamefont{Nishikawa}},
  \bibnamefont{et~al.}, \bibinfo{journal}{Phys.~Rev.~C}
  \textbf{\bibinfo{volume}{65}}, \bibinfo{pages}{034003}
  (\bibinfo{year}{2002}).

\bibitem[{\citenamefont{Ermisch et~al.}(2003)\citenamefont{Ermisch,
  Amir-Ahmadi, van~den Berg, Castelijns, Davids, Epelbaum, van Garderen,
  Gl\"ockle, Golak,  et~al.}}]{ermisch:03b}
\bibinfo{author}{\bibfnamefont{K.}~\bibnamefont{Ermisch}},
  \bibinfo{author}{\bibfnamefont{H.~R.} \bibnamefont{Amir-Ahmadi}},
  \bibinfo{author}{\bibfnamefont{A.~M.} \bibnamefont{van~den Berg}},
  \bibinfo{author}{\bibfnamefont{R.}~\bibnamefont{Castelijns}},
  \bibinfo{author}{\bibfnamefont{B.}~\bibnamefont{Davids}},
  \bibinfo{author}{\bibfnamefont{E.}~\bibnamefont{Epelbaum}},
  \bibinfo{author}{\bibfnamefont{E.}~\bibnamefont{van Garderen}},
  \bibinfo{author}{\bibfnamefont{W.}~\bibnamefont{Gl\"ockle}},
  \bibinfo{author}{\bibfnamefont{J.}~\bibnamefont{Golak}}, ,
  \bibnamefont{et~al.}, \bibinfo{journal}{Phys.~Rev.~C}
  \textbf{\bibinfo{volume}{68}}, \bibinfo{pages}{051001(R)}
  (\bibinfo{year}{2003}).

\bibitem[{\citenamefont{Ermisch et~al.}(2001)\citenamefont{Ermisch, van~den
  Berg, Bieber, Gl\"ockle, Golak, Hagemann, Hannen, Harakeh, de~Huu,
  Kalantar-Nayestanaki et~al.}}]{ermisch:01a}
\bibinfo{author}{\bibfnamefont{K.}~\bibnamefont{Ermisch}},
  \bibinfo{author}{\bibfnamefont{A.~M.} \bibnamefont{van~den Berg}},
  \bibinfo{author}{\bibfnamefont{R.}~\bibnamefont{Bieber}},
  \bibinfo{author}{\bibfnamefont{W.}~\bibnamefont{Gl\"ockle}},
  \bibinfo{author}{\bibfnamefont{J.}~\bibnamefont{Golak}},
  \bibinfo{author}{\bibfnamefont{M.}~\bibnamefont{Hagemann}},
  \bibinfo{author}{\bibfnamefont{V.~M.} \bibnamefont{Hannen}},
  \bibinfo{author}{\bibfnamefont{M.~N.} \bibnamefont{Harakeh}},
  \bibinfo{author}{\bibfnamefont{M.~A.} \bibnamefont{de~Huu}},
  \bibinfo{author}{\bibfnamefont{N.}~\bibnamefont{Kalantar-Nayestanaki}},
  \bibnamefont{et~al.}, \bibinfo{journal}{Phys.~Rev.~Lett.}
  \textbf{\bibinfo{volume}{86}}, \bibinfo{pages}{5862} (\bibinfo{year}{2001}).

\bibitem[{\citenamefont{Marcucci et~al.}()\citenamefont{Marcucci, Viviani,
  Schiavilla, Kievsky, and Rosati}}]{marcucci:05a}
\bibinfo{author}{\bibfnamefont{L.~E.} \bibnamefont{Marcucci}},
  \bibinfo{author}{\bibfnamefont{M.}~\bibnamefont{Viviani}},
  \bibinfo{author}{\bibfnamefont{R.}~\bibnamefont{Schiavilla}},
  \bibinfo{author}{\bibfnamefont{A.}~\bibnamefont{Kievsky}}, \bibnamefont{and}
  \bibinfo{author}{\bibfnamefont{S.}~\bibnamefont{Rosati}},
  \bibinfo{note}{nucl-th/0411083}.

\bibitem[{\citenamefont{Golak et~al.}(2000)\citenamefont{Golak, Kamada,
  Wita{\l}a, Gl\"ockle, Kuros, Skibi\'nski, Kotlyar, Sagara, and
  Akiyoshi}}]{golak:00a}
\bibinfo{author}{\bibfnamefont{J.}~\bibnamefont{Golak}},
  \bibinfo{author}{\bibfnamefont{H.}~\bibnamefont{Kamada}},
  \bibinfo{author}{\bibfnamefont{H.}~\bibnamefont{Wita{\l}a}},
  \bibinfo{author}{\bibfnamefont{W.}~\bibnamefont{Gl\"ockle}},
  \bibinfo{author}{\bibfnamefont{J.}~\bibnamefont{Kuros}},
  \bibinfo{author}{\bibfnamefont{R.}~\bibnamefont{Skibi\'nski}},
  \bibinfo{author}{\bibfnamefont{V.~V.} \bibnamefont{Kotlyar}},
  \bibinfo{author}{\bibfnamefont{K.}~\bibnamefont{Sagara}}, \bibnamefont{and}
  \bibinfo{author}{\bibfnamefont{H.}~\bibnamefont{Akiyoshi}},
  \bibinfo{journal}{Phys.~Rev.~C} \textbf{\bibinfo{volume}{62}},
  \bibinfo{pages}{054005} (\bibinfo{year}{2000}).

\bibitem[{\citenamefont{Schmid~{\it et al.}}(1996)}]{schmid:96a}
\bibinfo{author}{\bibfnamefont{G.~J.} \bibnamefont{Schmid~{\it et al.}}},
  \bibinfo{journal}{Phys.~Rev.~C} \textbf{\bibinfo{volume}{53}},
  \bibinfo{pages}{35} (\bibinfo{year}{1996}).

\bibitem[{\citenamefont{Smith and Knutson}(1999)}]{smith:99a}
\bibinfo{author}{\bibfnamefont{M.~K.} \bibnamefont{Smith}} \bibnamefont{and}
  \bibinfo{author}{\bibfnamefont{L.~D.} \bibnamefont{Knutson}},
  \bibinfo{journal}{Phys.~Rev.~Lett.} \textbf{\bibinfo{volume}{82}},
  \bibinfo{pages}{4591} (\bibinfo{year}{1999}).

\bibitem[{\citenamefont{Akiyoshi et~al.}(2001)\citenamefont{Akiyoshi, Sagara,
  Ueno, Nishimori, Fujita, Maeda, Nakamura, and Nakashima}}]{akiyoshi:01a}
\bibinfo{author}{\bibfnamefont{H.}~\bibnamefont{Akiyoshi}},
  \bibinfo{author}{\bibfnamefont{K.}~\bibnamefont{Sagara}},
  \bibinfo{author}{\bibfnamefont{S.}~\bibnamefont{Ueno}},
  \bibinfo{author}{\bibfnamefont{N.}~\bibnamefont{Nishimori}},
  \bibinfo{author}{\bibfnamefont{T.}~\bibnamefont{Fujita}},
  \bibinfo{author}{\bibfnamefont{K.}~\bibnamefont{Maeda}},
  \bibinfo{author}{\bibfnamefont{H.}~\bibnamefont{Nakamura}}, \bibnamefont{and}
  \bibinfo{author}{\bibfnamefont{T.}~\bibnamefont{Nakashima}},
  \bibinfo{journal}{Phys.~Rev.~C.} \textbf{\bibinfo{volume}{64}},
  \bibinfo{pages}{034001} (\bibinfo{year}{2001}).

\bibitem[{\citenamefont{Jourdan~{\it et al.}}(1985)}]{jourdan:85a}
\bibinfo{author}{\bibfnamefont{J.}~\bibnamefont{Jourdan~{\it et al.}}},
  \bibinfo{journal}{Phys.~Lett.} \textbf{\bibinfo{volume}{162B}},
  \bibinfo{pages}{269} (\bibinfo{year}{1985}).

\bibitem[{\citenamefont{Pitts~{\it et al.}}(1988)}]{pitts:88a}
\bibinfo{author}{\bibfnamefont{W.~K.} \bibnamefont{Pitts~{\it et al.}}},
  \bibinfo{journal}{Phys.~Rev.~C} \textbf{\bibinfo{volume}{37}},
  \bibinfo{pages}{1} (\bibinfo{year}{1988}).

\bibitem[{\citenamefont{Pickar et~al.}(1987)\citenamefont{Pickar, Karwowski,
  Brown, Hall, Hugi, Pollock, Cupps, Fatyga, and Bacher}}]{pickar:87a}
\bibinfo{author}{\bibfnamefont{M.~A.} \bibnamefont{Pickar}},
  \bibinfo{author}{\bibfnamefont{H.~J.} \bibnamefont{Karwowski}},
  \bibinfo{author}{\bibfnamefont{J.~D.} \bibnamefont{Brown}},
  \bibinfo{author}{\bibfnamefont{J.~R.} \bibnamefont{Hall}},
  \bibinfo{author}{\bibfnamefont{M.}~\bibnamefont{Hugi}},
  \bibinfo{author}{\bibfnamefont{R.~E.} \bibnamefont{Pollock}},
  \bibinfo{author}{\bibfnamefont{V.~R.} \bibnamefont{Cupps}},
  \bibinfo{author}{\bibfnamefont{M.}~\bibnamefont{Fatyga}}, \bibnamefont{and}
  \bibinfo{author}{\bibfnamefont{A.~D.} \bibnamefont{Bacher}},
  \bibinfo{journal}{Phys.~Rev.~C} \textbf{\bibinfo{volume}{35}},
  \bibinfo{pages}{37} (\bibinfo{year}{1987}).

\bibitem[{\citenamefont{Jans~{\it et al.}}(1987)}]{jans:87a}
\bibinfo{author}{\bibfnamefont{E.}~\bibnamefont{Jans~{\it et al.}}},
  \bibinfo{journal}{Nucl.~Phys.} \textbf{\bibinfo{volume}{A475}},
  \bibinfo{pages}{687} (\bibinfo{year}{1987}); 
  E. Jans (private communication).

\bibitem[{\citenamefont{Ishikawa et~al.}(1994)\citenamefont{Ishikawa, Kamada,
  Gl\"ockle, Golak, and Wita{\l}a}}]{ishikawa:94a}
\bibinfo{author}{\bibfnamefont{S.}~\bibnamefont{Ishikawa}},
  \bibinfo{author}{\bibfnamefont{H.}~\bibnamefont{Kamada}},
  \bibinfo{author}{\bibfnamefont{W.}~\bibnamefont{Gl\"ockle}},
  \bibinfo{author}{\bibfnamefont{J.}~\bibnamefont{Golak}}, \bibnamefont{and}
  \bibinfo{author}{\bibfnamefont{H.}~\bibnamefont{Wita{\l}a}},
  \bibinfo{journal}{Nuovo Cimento} \textbf{\bibinfo{volume}{A107}},
  \bibinfo{pages}{305} (\bibinfo{year}{1994}).

\end{thebibliography}

\end{document}